\documentclass[aps,prb,twocolumn,letterpaper]{revtex4}
\usepackage{graphicx}
\usepackage{dcolumn}
\usepackage{amsmath}
\usepackage{amssymb}
\usepackage{float}
\usepackage{epstopdf}
\usepackage{color}

\def\kv{{\bf k}}

\def\Sv{{\bf S}}

\def\beq{\begin{equation}}
\def\eeq{\end{equation}}

\begin{document}

\title{Strong Coupling Expansion of the Extended Hubbard Model with Spin-Orbit Coupling}
\author{Aaron Farrell and T. Pereg-Barnea}
\affiliation{Department of Physics and the Centre for Physics of Materials, McGill University, Montreal, Quebec,
Canada H3A 2T8}
\date{\today}
\begin{abstract}
We study the strong coupling limit of the extended Hubbard model in two dimensions. The
model consists of hopping, on-site interaction, nearest-neighbor interaction, spin-orbit coupling and Zeeman spin splitting.
While the study of this model is motivated by a search for topological phases and in particular a topological
superconductor, the methodology we develop may be useful for a variety of systems. We begin our treatment with a canonical transformation of the Hamiltonian to an effective model which is appropriate when the (quartic)
interaction terms are larger than the (quadratic) kinetic and spin-orbit coupling terms. We proceed by analyzing
the strong coupling model variationally. Since we are mostly interested in a superconducting phase we use a
Gutzwiller projected BCS wavefunction as our variational state. To continue analytically we employ the
Gutzwiller approximation and compare the calculated energy with Monte-Carlo calculations.  Finally we determine the topology of the ground state and map out the topology phase diagram.
  \end{abstract}
\maketitle

\section{Introduction}

First introduced in 1963\cite{Hubbard, Gutzwiller, Kanamori}, the Hubbard model is regarded by many to be the
simplest possible Hamiltonian which captures the essential physics of many-body systems with strong interactions. Owing to its simplicity, the Hubbard model has been used for decades to describe a
variety of systems. Its applications have ranged from antiferromagnetism\cite{anderson} to the
treatment of the metal-insulator transition\cite{Mott, Hubbard2}.

The standard Hubbard model contains hopping on the lattice sites (with hopping amplitude denoted by $t$) and on-site repulsion in the form of an energy penalty $U$ whenever two electrons are on the same site.  It has been studied extensively in two dimensions on the square lattice mainly in the context of the high T$_{\rm c}$ cuprates\cite{Scalapino1}.  The strongly interacting limit of this model at half filling is the Heisenberg Hamiltonian with antiferromagnetic coupling $J = 4t^2/U$.  Close to half filling (but not quite there) an appropriate approximation for the Hamiltonian is the t-J model where some hopping is allowed.  This model is challenging to deal with since it contains both quadratic (kinetic) terms and quartic (spin) terms.  Over the years many approaches where developed for the study of this strong interaction physics both analytically and numerically\cite{Lee}.  However, it is fair to say that at arbitrary doping $x$ (where $x$ is the hole density measured from half filling) any treatment uses some additional approximations.

In this paper we revisit the strong coupling limit of the Hubbard model with two additional terms (and hence the terminology `extended Hubbard').  The first is spin-orbit coupling which results in additional quadratic terms.  These can be regarded as spin dependent/spin flip hopping processes.   The second is off-site electron-electron interaction. We use this term to emulate the effect of the full Eliashberg method on the four fermion vertex\cite{Onari}.  In other words, instead of renormalizing the interaction vertex by the electron polarization bubble we add nearest neighbor attraction, $V<0$.  This effective attraction is appropriate whenever the polarization bubble of the fermions is maximal at $(\pi,\pi)$\cite{Scalapino}.

The model we consider here is motivated by recent interest in topological superconductors and their promise
to support Majorana fermions as bound excitations in vortex cores. The realization of
topological superconductors has been proposed in different setups such as semiconductor
heterostructures\cite{Fu,Sau,Alicea} and devices containing nanotubes\cite{lutchyn,oreg,Cook,klinovaja}.
In these heterostructures the topological superconductivity is a
result of spin-momentum locking provided by the spin-orbit coupling and tendency for pairing is induced
through proximity to a superconductor. Inspired by the above advances we have proposed a model in which
superconductivity arises from interactions (rather than proximity) in a system with spin-orbit
coupling\cite{Farrell}.

While the above model was studied in the weak coupling limit\cite{Farrell}, superconductivity occurs at
intermediate to strong coupling. In this Paper we lay the foundation of a strong coupling study of the extended
Hubbard model in the presence of off-site interaction as well as spin-orbit coupling. We believe that this
versatile model is useful beyond the scope of our specific application.  For example, it might be useful in describing complex oxide heterostructure interfaces where the inversion symmetry is broken due to the superlattice.  The broken inversion symmetry gives rise to Rashba spin-orbit coupling at the interface. {Further,  we feel that the methods developed here may be of use in other applications. One example of this might be the Kane-Mele-Hubbard model, which,  among other things, is relevant to studies of how interactions effect topological band structures\cite{rachel1, rachel2}}. It is therefore desirable to get a handle on the
strong coupling limit of our model and this is the purpose of the present work.

To achieve the strong coupling limit of our model we assume that the interaction part of
the Hamiltonian is large compared to the kinetic and spin-orbit coupling terms and develop a strong-coupling
expansion. When this expansion is truncated at the second order and specialized to electron densities close to half filling it is a generalization of the $t-J$ model with Dzyaloshinskii-Moriya and compass anisotropy in the spin interaction.  In this regime we employ the Gutzwiller approximation\cite{Gutzwiller, zhang, zhangprl, vollhardt} to study the resulting model. This approximation has been successful in describing d-wave pairing and superconducting phase fluctuations\cite{Paramekanti1,Paramekanti2} in the standard Hubbard model in two dimensions.

The Gutzwiller approximation consists of two stages. In the first a variational wavefunction is generated. When pairing is present, the appropriate
variational wavefunction is a Gutzwiller projected BCS wavefunction\cite{Paramekanti1, , sorella1, sorella2}. This wavefunction
may contain any mean-field like orders (such as density waves, superconductivity or antiferromagnetism) and the
Gutzwiller projection builds the strong interactions into it by eliminating any doubly occupied configurations.
At this point one should evaluate the variational energy as the expectation value of the Hamiltonian with respect to the projected mean field wavefunction and minimize it with respect to the mean field order parameters.

The minimization of the variational energy in the many-body projected state is not trivial due to the projection
operation and can not be done analytically. The evaluation should be done repeatedly until the energy is
minimized and this can be done numerically using the Monte-Carlo technique\cite{future_work}. In the current
work we choose to take a different approach and proceed analytically by making another approximation. Assuming
that the charge (holes) is distributed uniformly on the lattice we may estimate the effect of the Gutzwiller
projection on the various components of the variational energy. We may therefore renormalize the Hamiltonian
parameters instead of performing the projection. This is the second stage of the Gutzwiller approximation which leads to a mean-field-like Hamiltonian whose parameters depend on the filling and can be analyzed in the standard way.

The rest of this article is organized as follows. In the next section we introduce the model and briefly present the generalized strong coupling expansion while leaving the details to Appendix~\ref{ap:expansion}. Once we have obtained the effective Hamiltonian, we
project it on a sub-space which is close to half-filling on the hole doped side and obtain a generalization of the $t-J$ model.
We then study the strong coupling Hamiltonian first without spin-orbit coupling and a Zeeman term in section~\ref{sec:Gutwiller}, and then with finite spin-orbit coupling in section~\ref{sec:SOC}.  We conclude with a mapping of the superconductor's phase diagram according to the ground state topology.

\section{Strong Coupling Expansion}\label{sec:SCE}

\subsection{The Model}
The extended Hubbard model we consider is given by the following Hamiltonian on a two-dimensional square lattice:
\begin{equation}
H=T+H_{SO}+H_{int}.
\end{equation}
where
\begin{equation}
T = -\frac{t}{2}\sum_{i,\delta,\sigma} (c_{i,\sigma}^\dagger c_{i+\delta,\sigma}+ c_{i+\delta,\sigma}^\dagger
c_{i,\sigma})
\end{equation}
is the tight binding kinetic energy and
\begin{equation}
H_{SO} = \sum_{\kv} \psi_{\kv}^\dagger \mathcal{H}_\kv \psi_{\kv},
\end{equation}
where  $\psi_{\kv}=(c_{\kv,\uparrow}, c_{\kv,\downarrow})^T$, $\mathcal{H}_\kv = {\bf d}_\kv \cdot \vec{ \sigma}$ (with $\vec\sigma$ a vector of Pauli matrices acting on the spin) and ${\bf d}=(A\sin{k_x},A\sin{k_y},2B(\cos{k_x}+\cos{k_y}-2)+M)$ ($A,B$ and $M$ are material parameters which describe the various spin-orbit coupling and Zeeman strengths). The three parameters in the
spin-orbit coupling model above may originate from a variety of different sources. For example the
parameters $A,B$ may come from a traditional spin-orbit model like the Rashba and Dresselhaus terms in Refs.
[\onlinecite{Sau},\onlinecite{Alicea}]. A second source may be parameters such as those used in the BHZ
model, suitable for quantum wells\cite{rothe, lu, guigou}. Similarly, the Zeeman field parameter $M$ may be the result of a band gap\cite{BHZ}, an external magnetic field or a magnetic field of a nearby ferromagnetic layer\cite{Sau}. As $M$ may come from a
variety of sources we will ignore any orbital effects that could arise in some cases.

The interaction terms denoted by $H_{int}$ above are given by
 \begin{equation}
H_{int} = U \sum_i n_{i\uparrow} n_{i,\downarrow}+ V\sum_{\langle i,j\rangle} n_{i} n_{j},
\end{equation}
where $U>0$ is the on-site repulsion strength and $V<0$ describes attraction between nearest neighbors.

For the purpose of making the strong coupling expansion we will need to express the Hamiltonian in real space. We therefore transform the spin orbit coupling to real space and divide it into two types of terms.  $H_z$, the Zeeman Hamiltonian will contain all on-site terms, proportional to $\sigma_z$:
 \begin{equation}
 H_Z = \sum_{\alpha, i} \sigma^z_{\alpha\alpha} (M-4B)c^\dagger _{i,\alpha} c _{i,\alpha}.
 \end{equation}
 The other terms in $H_{SO}$ amount to hopping processes which act non-trivially on the spin.  We write them in real space and combine them with the hopping into a $2\times2$ generalized hopping matrix such that:
\begin{eqnarray}\label{tgen}
T+H_{SO} &=& H_z + T+H_{SO} = \sum_{i,\delta,\alpha,\beta}
\hat{t}_{\alpha\beta}(\vec{\delta}) c^\dagger_{i,\alpha}c^{}_{i+\delta,\beta}+H_Z \nonumber \\
\hat{t} &=& \begin{pmatrix}-t +B & -i{A\over2}\delta_x - {A\over 2}\delta_y \\-i{A\over 2}\delta_x + {A\over2}\delta_y & -t-B\end{pmatrix}
\end{eqnarray}

The reader should note that this model has been proposed not with a specific physical system in mind but for the sake of versatility. The work to follow is meant to motivate a search for a specific material with these properties. One possible candidate that may be described by this model is  copper intercalated Bi$_2$Se$_3$ which, although a 3D material, develops a 2D-like Fermi surface at certain doping\cite{ lahoud}.

\subsection{The strong coupling Hamiltonian}
The strong coupling expansion distinguishes between the high interaction energy scales $U,V$ and other, quadratic terms in the Hamiltonian.  Written in real space the quadratic terms are either on-site (chemical potential, Zeeman) or hopping.  The on-site terms do not change the potential energy while the hopping terms do.  Ideally, one would like to diagonalize the Hamiltonian but due to its quartic terms this can not be done analytically.  Instead of diagonalizing we set out to block diagonalize the Hamiltonian. Since we are interested in the strong coupling regime we would like to block-diagonalize the Hamiltonian such that the interaction energy is constant at each block and work at the lowest interaction energy block.  In other words - there exist a unitary transformation which eliminates terms which change the interaction energy from the Hamiltonian.  Using the properties of the desired transformation (unitarity and interaction energy conservation) we can formally write down the transformation.  In order to find a closed form, however, we resort to a power expansion in a small parameter of the order of the ratio between quadratic part of the energy and the interaction part of it.  We are then able to find the transformation and the transformed Hamiltonian up to a given order.

 \begin{figure}[tb]
  \setlength{\unitlength}{1mm}

   \includegraphics[scale=.4]{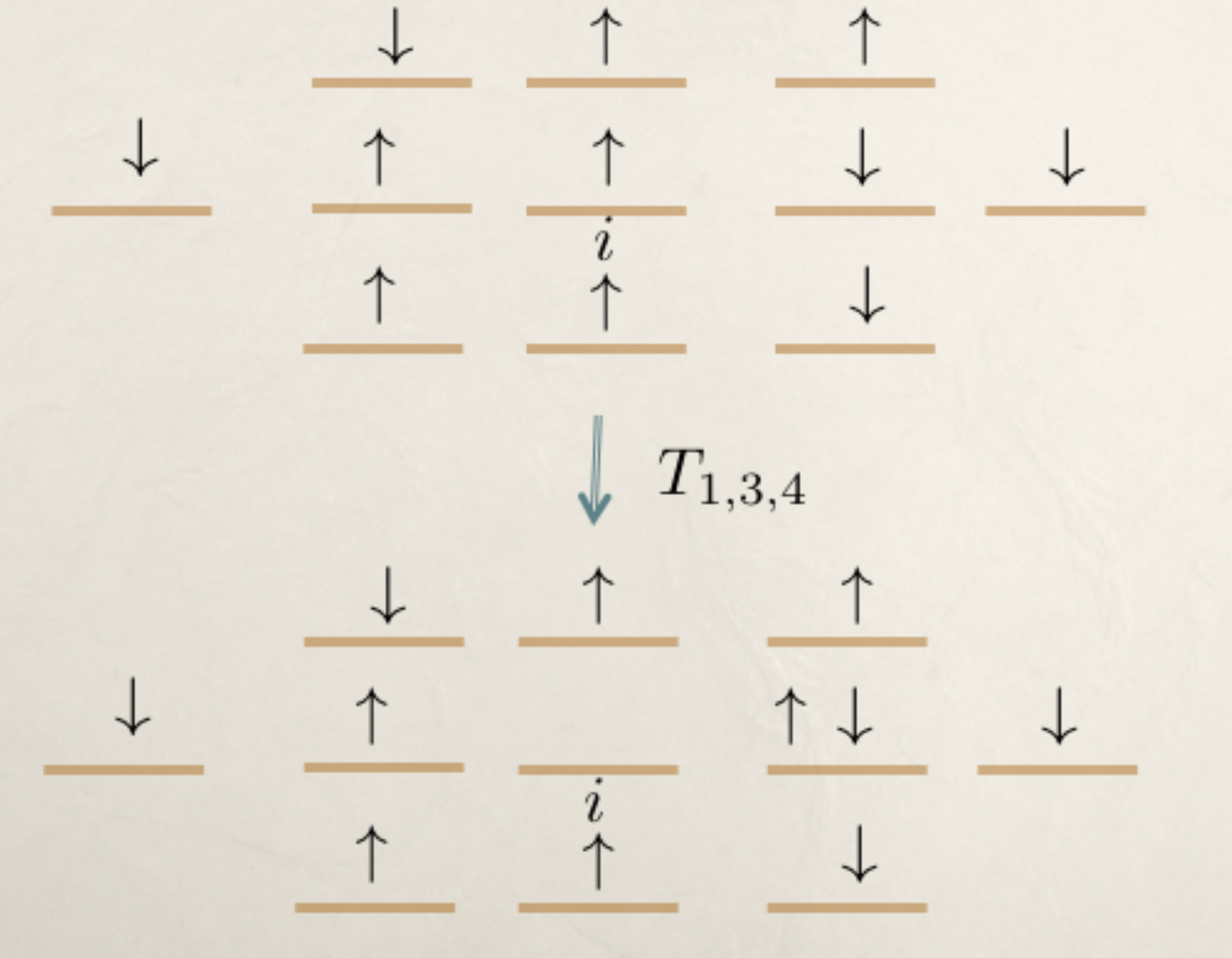}
\caption{{\small
Schematic diagram of the a hopping process contributing to $T_{1,3,4}$.
     }
     }\label{fig:hopping}
\end{figure}

Let us sketch the procedure which resembles the treatment of ref's. [\onlinecite{macdonald}] of the original Hubbard model.  We denote the unitary transformation that block diagonalizes the Hamiltonian by $\exp(iS)$ and the Hamiltonian terms excluding the (constant) interaction  and Zeeman energy by  $H_1= T+H_{SO}$. In order to perform a series expansion for both the unitary transformation and the transformed Hamiltonian we arrange the terms in $H_1$ by their effect on the interaction energy.  For example, the operator $T_0$ includes all terms which do not change the interaction energy at all.  We further define the operators $T_{m,N_2,N_1}$ as a collection of the hopping terms which change the number of doubly occupied sites by $m$ and the number of nearest neighbor pairs the electron sees from $N_1$ to $N_2$.  This amounts to a change of $mU+(N_2-N_1)V$ to the interaction energy.  As an example of one of these processes we refer the reader to Fig. \ref{fig:hopping}. In the figure we illustrate one of the hopping processes involved in the specific operator $T_{1,3,4}$. An electron starts on lattice site $i$ and hops to a nearest neighbour. {In the process of this hop we see a double occupancy is created and so $m=1$. Further, the electron that we are hopping begins on lattice site $i$ which has 4 nearest neighbours and then hops (to the right in the figure) to a site with 3 nearest neighbours. Thus this process has $N_1=4$ and $N_2=3$}  After a few more steps we arrive at the expansion and present it up to second order:
\begin{eqnarray}\label{strongH}
\tilde {H} &=& H_{int} + H_Z + \sum_{M} \ {T}_{0,M,M} \\ \nonumber &+& \sum_{m,M_1,N_1,M_2}' \frac{ {T}_{m,M_1,
N_1}  {T}_{-m, M_2, M_1+M_2-N_1}}{(mU+(M_1-N_1)V)}+\mathcal{O}(1/U^2)\nonumber \\
&=&H_{int}+H_1,
\end{eqnarray}
where the primed sum excludes terms in which $mU+(M_1-N_1)V=0$.

 The results presented in Eq. (\ref{strongH}) are general for any $U,|V| \gg t, A, B$.  Moreover, since the denominator of the above expression is a combination of $U$ and $V$ it is sufficient to have only one of them large compared with the band width.  In our system one may expect that the on-site repulsion $U$ be large while the nearest neighbor attraction is small.  Nevertheless, our formalism allows us to treat both interaction terms exactly and we do so.

 To proceed we focus on the parameter regime $U\gg |V|$ (and $V$ is actually smaller than $t$ such that it does not induce any charge clustering). For this parameter regime the ground state of a system on the hole-doped side of half-filling before we account for $T$ or $H_{SO}$ will lie in the subspace of real-space configurations with no doubly occupied sites\cite{macdonald, dongen}. We consider a complete subspace of no doubly occupied sites even though the effects of $V$ may lead so some phase separation of electrons and holes. We feel this to be an appropriate simplification to make in order to make progress with the problem.  At half filling the second part of the Hamiltonian that is second order in the $T_{m,M_1,M_2}$ operators can be rearranged as a spin Hamiltonian which has been studied elsewhere\cite{SpinHamiltonian}. At, or slightly {\em hole} doped away from, half-filling the dominant contribution from the term of the form ${T}_{m,M_1,N_1}  {T}_{-m, M_2, M_1+M_2-N_1}$ will see an initial number of nearest-neighbours $M_1+M_2-N_1=4$ and hop to a site with $M_2=3$. All other terms involved in the second order contribution are either impossible or occur with a probability proportional to some power of $x$, which is very small near half-filling. Thus although other terms that are of order $1/U$ exist in the second order term in Eq. (\ref{strongH}),  the term with denominator $ -U+V$ is the {\em statistically} dominant term.  A picture of this process can be formed by looking at Fig. \ref{fig:hopping}. This process corresponds to ${T}_{-1,4,3}  {T}_{1, 3, 4}$ which corresponds to the hop shown in Fig. \ref{fig:hopping} and then one of the two electrons on the site to the right of site $i$ hopping back.    Close to, but not at half filling hopping without a change to the interaction energy is possible and included in $T_{0,M,M}$.  This is the generalization of the well known $t-J$ model to our case and the subject of this paper.  The model is given by:
\begin{equation}\label{tjd}
H_{t,J_\delta} = H_Z + \sum_{N} T_{0,N,N} +\sum_{i,\delta}J_\delta^{\mu,\nu} \Sv_i^\mu \Sv_{i+\delta}^\nu.
\end{equation}
where the spin vector is defined by $S_i^\mu = \frac{1}{2} \sum_{\alpha,\beta} c^\dagger_{i,\alpha} \sigma^\mu_{\alpha,\beta} c_{i,\beta}$ and the spin coupling is given by
\begin{eqnarray}
&& {\cal J}_{\delta}={1\over 2(U-V)} \times \\ \nonumber
  &&\small{\left(\begin{matrix}
      4t^2+ A^2a(\delta)-4B^2 & 0& -4At\delta_y \\
      0 & 4t^2- A^2a(\delta)-4B^2&4At\delta_x \\
      4At\delta_y&-4At\delta_x&4t^2-A^2+4B^2\\
   \end{matrix}\right)}
\end{eqnarray}
where $\delta$ is the nearest neighbor vector and $a(\delta)$ is $1$ for $\delta = \pm \hat x$ and $-1$ for $\delta = \pm \hat y$.
From the above matrix one can read off the Heisenberg, Dzyaloshinskii-Moriya and compass anisotropy terms:
\begin{eqnarray}
&&\sum_{i,\delta}{\cal J}_\delta^{\mu,\nu}  \Sv_{i}^\mu\Sv_{i+\delta}^\nu = \sum_{i,\delta}\big{[} J\Sv_i\cdot\Sv_{i+\delta} + D\delta\cdot\left(\Sv_i\times\Sv_{i+\delta}\right) \nonumber \\ && + C(\delta)\Sv_i^x\Sv_{i+\delta}^x+ C'(\delta)\Sv_i^y\Sv_{i+\delta}^y\big{]} \nonumber \\
&&J = {4t^2-A^2+4B^2\over 2(U-V)} \nonumber \\
&&D ={2At \over U-V} \nonumber \\
&&C(\delta) = \frac{A^2+A^2a(\delta)-8B^2}{2(U-V)} \;\;\;\;  \\ \nonumber && C'(\delta) =  \frac{A^2-A^2a(\delta)-8B^2}{2(U-V)}
\end{eqnarray}
Spin models similar to the one above have been employed in the study of spin-orbit effects in ultra-cold atoms and other systems\cite{Banerjee, Radic, Cole}.
The focus of the remainder of this paper will be on studying a renormalized version of the above model.

\section{A Gutzwiller Approximation at Strong Coupling}\label{sec:Gutwiller}
In order to find the ground state of the strong coupling limit of our model we adapt the Gutzwiller approximation\cite{zhang, zhangprl, vollhardt} to the extended Hubbard model.
The rational of this approximation is as follows.  The Gutzwiller approximation is a variational method that uses a projected BCS mean-field wavefunction.

A projected BCS wavefunction has proved useful in the context of strongly correlated electron systems.
It is obtained through projecting out all doubly occupied configurations from the BCS wavefunction.  The resulting wavefunction,
$P_G|\psi_{MF}\rangle$ encodes both the tendency for developing order such as Cooper pairing and density waves due to its mean field starting point as well as the strong on-site repulsion of the model due to the Gutzwiller projection. The reader should note that more sophisticated generalizations of this wave function have been used in Monte Carlo calculations in the past\cite{sorella3, sorella4}

The variational energy $E_v=\langle \psi_{MF}|P_G HP_G|\psi_{MF}\rangle$ should be calculated and minimized with respect to the order parameters built into $|\psi_{MF}\rangle$.  However, the application of the Gutzwiller projection on the wavefunction makes the evaluation very complicated and not attainable analytically. One may then resort to numerical methods such as Monte Carlo integration in order to find $E_v$\cite{Paramekanti1, sorella1, sorella2}.  Monte Carlo methods have been refined in recent years to the point that they can be regarded as  variationally  exact.  We take this numerical
approach elsewhere\cite{future_work}.  Here we employ an analytical approach, namely, the Gutzwiller approximation.
The Gutzwiller approximation which was developed several decades ago\cite{zhang, zhangprl, vollhardt} is a method to approximately apply the Gutzwiller projection operator. First, one should note that the projection operator can be applied either on the wavefunction or on both sides of the Hamiltonian.  This means that one may derive an effective Hamiltonian that acts on the non-projected mean-field wavefunction.  The resulting energy is therefore an approximation of the expectation value of the original Hamiltonian with respect to the projected wavefunction.  In this section we apply the Gutzwiller approximation to the extended Hubbard model, starting without spin-orbit coupling and then adding it in the next section.

Without spin orbit coupling and Zeeman field Eq.~(\ref{tjd}) reduces to:
\begin{eqnarray}
H_{t,J} = \sum_{N} T_{0,N,N} +\sum_{i,\delta}J  \Sv_{i}\cdot \Sv_{i+\delta}.
\end{eqnarray}
where the only difference from the usual $t-J$ model is in the kinetic term.  In this term we only include hopping processes which do not change the number of doubly occupied sites as well as the number of nearest neighbor pairs.
We assume  hole doping, $x$ in the vicinity of half filling such that $\langle n \rangle =1-x$ and $x\ge0$. We further assume that the system is
completely unpolarized, {\it i.e.} $\langle n_{\uparrow}\rangle =\langle n_{\downarrow} \rangle = \frac{1}{2}\langle n\rangle$.  While this may be expected
for a spin conserving Hamiltonian, it may not be locally accurate for a system with spin-orbit coupling and certainly will not be favorable when a Zeeman term is included.  The variational energy is given by:
\begin{equation}
E_{var}= \frac{\langle \psi_{MF} | P_G H_{t,J} P_G|\psi_{MF}\rangle}{\langle \psi_{MF}
|P_G|\psi_{MF}\rangle}.
\end{equation}
Note that since our projection will be applied to the Hamiltonian rather than the wavefunction, we do not need to specify the mean-field wavefunction just yet.

First let us estimate the change in the the expectation value of the hopping term due to the Gutzwiller projection and the definition of $T_{0,N,N}$ which is restricted to hopping processes which do not change the interaction energy.  The kinetic part of the variational energy is therefore
 \begin{widetext}
 \begin{eqnarray}\label{Kgut}
 K_G &=& \sum_N \frac{\langle \psi_{BCS} | P_G  T_{0,N,N}P_G|\psi_{BCS}\rangle}{\langle \psi_{BCS}
 |P_G|\psi_{BCS}\rangle}
 =\sum_{N,i,\delta,\alpha} \sum_{\Sigma[\mathcal{N}_1], \Sigma[\mathcal{N}_2]=N} \frac{\langle \psi_{BCS} | P_G O_{i+\delta}[\mathcal{N}_2]
 c^\dagger_{i+\delta, \alpha}c_{i, \alpha}O_{i}[\mathcal{N}_1]P_G|\psi_{BCS}\rangle}{\langle \psi_{BCS}
 |P_G|\psi_{BCS}\rangle}
 \end{eqnarray}
 \end{widetext}
 where the operator $O_i[\mathcal{N}]$ projects out all possible spin orientations of the nearest neighbors of site $i$ except the one specified by $\mathcal{N} = (\mathcal{N}_{x\uparrow},\mathcal{N}_{x\downarrow},\mathcal{N}_{-x\uparrow},\mathcal{N}_{-x\downarrow},
 \mathcal{N}_{y\uparrow},\mathcal{N}_{y\downarrow},\mathcal{N}_{-y\uparrow},\mathcal{N}_{-y\downarrow})$, where $\mathcal{N}_{\delta\sigma}$ takes a value of 1 (a value of zero) for site $i+\delta$ occupied (unoccupied) by an electron with spin $\alpha$. A formal definition of this operator is left to Appendix~\ref{ap:expansion}.  Let us consider the action of $ P_G O_{i+\delta}[n_2]
 c^\dagger_{i+\delta, \alpha}c_{i, \alpha}O_{i}[n_1]P_G$ on any configuration that may appear in the BCS state.  First $P_G$ removes all possible double occupancies. For a system close to half filling this projection means that the only configurations left in
$|\psi_{BCS}\rangle$ are ones with singly occupied or unoccupied sites.
$T_{0,N,N}$ then moves an electron from an occupied site with $N$ nearest neighbors to an empty site with $N$
nearest neighbors. Let us compare this to the normal unprojected operator, $c^\dagger_{i+\delta,
\alpha}c_{i, \alpha}$, acting on just the BCS ground state. When this term acts, it finds a lattice site which
is occupied by an electron of some spin and then moves it to a neighbor site that is not occupied by this
particular spin.

We can assign relative probabilities for each of these two events. For the latter case the probability that the process occurs is just the probability that
the lattice site the electron moves to is void of the particular spin we are moving. Assuming that the
probability that any site is occupied is uniform then the probability of the particular move we just described is $1-\langle n_{i,\alpha}\rangle=
(1+x)/2$. For the case of the projected expectation value there are several concurrent requirements we must
meet. First, we require the lattice site where the electron is moved to be empty.  This gives a factor of $x$. Second, the site we move the electron from must have exactly $N$ occupied nearest
neighbors sites. The site we are moving the electron to is a nearest neighbor site and must be empty, therefore $N\le3$. The probability of having $N\le3$
occupied nearest neighbor sites is then the binomial probability of having exactly $N$ successes in 3 trials
with success probability $1-x$, in other words $\binom{3}{N}(1-x)^Nx^{3-N}$. Finally, the site we move to must
also have exactly $N$ occupied nearest neighbors, this again occurs with probability
$\binom{3}{N}(1-x)^Nx^{3-N}$. Combining these three observations we find that the probability of the entire
series of events is $x(\binom{3}{N}(1-x)^Nx^{3-N})^2$.

This means that
 \begin{widetext}
\begin{equation}
 \frac{\langle \psi_{BCS} | P_G O_{i+\delta}[n_2] c^\dagger_{i+\delta, \alpha}c_{i,
  \alpha}O_{i}[n_1]P_G|\psi_{BCS}\rangle}{\langle \psi_{BCS}  |P_G|\psi_{BCS}\rangle} \simeq
 \frac{2x}{1+x}\left(\binom{3}{N}(1-x)^Nx^{3-N}\right)^2  \frac{\langle \psi_{BCS} |  c^\dagger_{i+\delta,
  \alpha}c_{i,  \alpha}|\psi_{BCS}\rangle}{\langle \psi_{BCS}  |\psi_{BCS}\rangle}
\end{equation}
 \end{widetext}
and we therefore redefine
\begin{eqnarray}
t \to g_t t = \frac{2xt}{1+x}
\sum_{N=0}^3\left(\binom{3}{N}(1-x)^Nx^{3-N}\right)^2
\end{eqnarray}
We now turn to the quartic terms.  Although our goal is to obtain something that is written as a spin-spin interaction we consider the double hopping processes that give rise to them as this language is more suitable for the projection.  The interaction energy can be written as the expectation value of the quartic terms in $\tilde{H}$ and reads
  \begin{widetext}
\begin{equation}\label{defVg}
V_G= \sum_{m,M_1,N_1,M_2}' \frac{ 1}{(mU+(M_1-N_1)V)}  \frac{\langle \psi_{BCS} | P_G {T}_{m,M_1, N_1}  {T}_{-m,
M_2, M_1+M_2-N_1}P_G|\psi_{BCS}\rangle}{\langle \psi_{BCS}  |P_G|\psi_{BCS}\rangle}
\end{equation}
 \end{widetext}
where the primed sum is as defined before.

First we repeat that in systems with hole doping
$x\ge0$ and after $P_G$ is applied to the wavefunction, there are only singly occupied or empty sites left in each configuration in the mean field state. This means that $-m$ can never be $-1$, as it is impossible to reduce the
number of doubly occupied sites. Therefore $-m$ is either $1$ or $0$. We will ignore the $m=0$ terms above, these terms require the movement of two electrons to
two empty sites and so are a factor of $x^2$ less probable than their $m=-1$ counterparts. Let us then consider only ${T}_{-1,M_1, N_1}  {T}_{1, M_2,M_1+M_2-N_1}$. This term creates a doubly occupied site and then destroys it.  This can happen in two ways: (1) an electron hops to an occupied site and then one of these two electrons hops
back to the original site or, (2) an electron hops to an occupied nearest neighbor and then one of the two
electrons on this site hops to an empty next-nearest neighbor site. Keeping with our motive of simplicity,
we will ignore the latter ``three site" move as it is less probable near half filling. Taking all of these concerns into account we have

\begin{equation}\label{defVg2}
V_G= \sum_{M_1,N_1} \frac{\langle \psi_{BCS} | P_G {T}_{-1,M_1, N_1}  {T}_{1 N_1,
M_1}P_G|\psi_{BCS}\rangle}{((M_1-N_1)V-U)\langle \psi_{BCS}  |P_G|\psi_{BCS}\rangle}
\end{equation}
We now estimate the Gutzwiller renormalization for $V_G$ . First consider a non-projected process. In this case an electron hops to an occupied site and then one of the two electrons at the site hops back. In order for
this to occur both sites must only be occupied by one spin. The probability of having a site only occupied by (say) an up spin next to a site which is only occupied by a
down spin is $\left(\frac{1-x}{2}\right)\left(\frac{1+x}{2}\right)\left(\frac{1-x}{2}\right)
\left(\frac{1+x}{2}\right)$. Now we consider the probability with all of the appropriate projections. The net
result of $P_G {T}_{-1,M_1, N_1}  {T}_{1 N_1, M_1}P_G$ is that an electron with an up spin hops from a site with $M_1$ nearest neighbors to a site with which is occupied by a down spin with $N_1$ nearest neighbors and then one of the two electrons now occupying this nearest neighbour site hops back to the original site.
$M_1$ can range from $1$ to $4$ while $N_1$ can take values from $0$ to $3$. The
probability of this process is $\left(\frac{1-x}{2}\right)^2 \binom{3}{N_1}(1-x)^{N_1}
x^{3-N_1}\binom{4}{M_1}(1-x)^{M_1}x^{4-M_1}$. We therefore approximate the interaction energy by
\begin{eqnarray}\label{defVg3}
V_G&\simeq& -\tilde{J}\sum_{i,\delta, \alpha, \alpha'} {\langle \psi_{BCS} | c_{i+\delta,  \alpha}^\dagger
c_{i, \alpha} c_{i, \alpha'}^\dagger c_{i+\delta, \alpha'}|\psi_{BCS}\rangle} \nonumber \\
&\equiv& \tilde{J}\sum_{i,\delta}\langle  \Sv_i\cdot\Sv_{i+\delta} \rangle\nonumber \\
\tilde{J} &=& g_J g_U\nonumber \\
g_J &=& \frac{4t^2}{(1+x)^2}  \\ \nonumber
g_U&=&\sum_{N_1=0}^3 \sum_{ M_1=1}^4 \frac{(1-x)^{N_1}
 x^{3-N_1}(1-x)^{M_1}x^{4-M_1}}{U-(M_1-N_1)V}  \binom{3}{N_1}\binom{4}{M_1}.
 \end{eqnarray}
where we have omitted the quadratic terms which originate from rearranging the fermion operators.

The next step in the Gutzwiller approximation is to evaluate the variational energy (this time with a non-projected BCS wavefunction) and minimize it with respect to the order parameters.  As this step is similar to the standard variational mean field procedure we de not give the details here.  Some sample results and their comparison to a numerical Monte-Carlo evaluation are given in Appendix~\ref{ap:NoSOC-BCS}.

\section{Finite Spin-Orbit Coupling} \label{sec:SOC}
In this section we keep the parameters
$A,B$ and $M$ finite. Keeping either $M$ or $B$ finite leads to spin-polarization of the electrons in the system. This inequality in the number of spin-up and spin-down electrons complicates the nature of the Gutzwiller approximation as the tacit assumption we made in the last
section that $\langle n_{\uparrow}\rangle   =\langle n_{\downarrow}\rangle   =\langle n \rangle/2$ is now
violated. We therefore must estimate the Gutzwiller renormalization for processes involving different spin
flavors separately.

We begin by considering the hopping term $T_{0,M,M}$ in Eq.(\ref{strongH}) and estimating the effect of the Gutzwiller projection. The first projection we deal with is the restriction on the number of nearest neighbours permitted before and after an electron hops from one site to the other. As this part of the projection depends only on the site occupation and not the spin it results in a factor similar to the one we had before:
\beq
g_{NN} = \sum_{N=0}^3\left(\binom{3}{N}(1-x)^Nx^{3-N}\right)^2
\eeq

Next we treat the different parts of $T_{0,M,M}$ separately. Recall that we defined the generalized hopping
matrix, $\hat{t}$, in Eq.~(\ref{tgen}) as having two parts. One includes simple hopping (which may have different amplitudes for different spins) and the other includes processes in which the spin is flipped during the hopping.  We estimate the effect of the projection on each one of the processes by considering all possible ways this process can occur in some state $|\psi\rangle$ and all possible ways they can occur in a Gutzwiller
projected state $ P_G |\psi\rangle$\cite{zhang, Ko, Chou}.  This leads to the renormalized hopping matrix.
\begin{eqnarray}
&&g_t( \alpha) = \frac{x}{1-\langle n_{ \alpha} \rangle } \\ \nonumber &&g_A= \frac{x}{\sqrt{(1-\langle
n_{\uparrow} \rangle)(1-\langle n_{\downarrow} \rangle)} } \\ \nonumber
\hat{t}^{g}_{\alpha,\alpha'}( \delta)   &=& (B \sigma^z_{\alpha,\alpha} -t)g_t( \alpha) g_{NN}\delta_{ \alpha, \alpha'}\\ \nonumber&-&g_A
g_{NN}\frac{Ai}{2} (\delta_x+\sigma^y_{\alpha, \bar{\alpha}}\delta_y)  \delta_{ \alpha',\bar{ \alpha}}
\end{eqnarray}
We can see that $g_t( \alpha)$ is the relative probability of a hop of an electron with spin $\alpha$. In the projected term we require an empty site in order to make the hop while in the unprojected term we
simply need the destination site to not be occupied by a spin $\alpha$. Similarly we can understand $g_A$.  The projected process requires an empty site and hence the factor of
$x$ in the numerator. On the other hand, the unprojected process requires that the site the electron hops to has no electron with the opposite spin. This accounts for a factor $1- \langle n_{ \alpha}
\rangle$, however we must equally weight this with the probability for the process to happen in reverse, this is
where the factor of $1- \langle n_{\bar{ \alpha}} \rangle$ and the square root come from.

To obtain the Gutzwiller factor of the quartic term in our strong coupling model at finite spin-orbit coupling we (again) neglect three
site hopping and processes where the $m$ (in the $T_{m,N_1,N_2}$ operators) is zero. This leaves us with only
 terms that begin and end on the same site which can be written as: $\sum_{i,\delta} \tilde{J}_{\delta}^{\mu\nu} \Sv_{i}^\mu\Sv_{i+\delta}^\nu$.
Let us rewrite the coupling matrix as:
\begin{eqnarray}\label{JGA1}
\small{\small{ \tilde{J}_{\delta}=\frac{1}{2} \left(\begin{matrix}
      J_1+J_2a(\delta)-J_4 & 0& -J_3\hat{y}\cdot \vec{\delta} \\
      0 & J_1-J_2a(\delta)-J_4&J_3\hat{x}\cdot \vec{\delta} \\
      J_3\hat{y}\cdot \vec{\delta}&-J_3\hat{x}\cdot \vec{\delta}&J_1-J_2+J_4\\
         \end{matrix}\right)}}
\end{eqnarray}
where we have defined the three couplings $J_1=4g_Ut^2$,  $J_2=g_UA^2$,  $J_3=4g_UAt$ and
$J_4 = 4g_UB^2$ and the function $a(\delta)$ is 1 if $\delta=\pm \hat{x}$ and -1 if $\delta=\pm\hat{y}$.
Here we have already approximately taken the restrictions on the number of nearest neighbors permitted by a
given hop by replacing the denominator by the Gutzwiller renormalization $g_U$ which has been defined in Eq.~(\ref{defVg3}).
We have not yet taken the occupancy requirements into account.  We now consider the factors which are spin dependent to obtain individual renormalization factors for different matrix elements.

 To carry out this renormalization one can group the spin Hamiltonian $\sum_{i,\delta} \tilde{J}_{\delta}^{\mu\nu} \Sv_{i}^\mu\Sv_{i+\delta}^\nu$ into four separate terms, $H_{J,\text{i}}$ through $H_{J, \text{iv}}$. Each one of these four terms describes a different physical process. The first term, $H_{J,1}$, flips the spins on nearest neighbour sites ({\it e.g. } an up spin on site $i$ and a down spin on site $j$ are flipped to a down and an up spin respectively). The second term looks at the $z$ component of the spins on neighbouring lattice sites and takes the form $S^z_iS^z_j$. The third term looks for nearest neighbour sites with two up or two down spins and flips both spins. Finally, the fourth flips a spin on site $j$ while measuring the $z$ component of a spin on site $i$. By considering the amplitude for each of these four physical processes in the projected and unprojected state, we can form Gutzwiller renormalization factors for each of them. We abdicate the details of this process to Appendix \ref{sec:Gfactors},  the result of performing this weighting of amplitudes is as follows
  \begin{eqnarray}
  g_{J,\text{i}} &=& \frac{1}{(1-\langle n_{\uparrow} \rangle)(1-\langle n_{\downarrow}\rangle)} \\ \nonumber
  g_{J,\text{ii}} &=& 1 \\ \nonumber
    g_{J,\text{iii}} &=&  \frac{1}{(1-\langle n_{\uparrow}\rangle)(1-\langle n_{\downarrow}\rangle)} \\ \nonumber
      g_{J,\text{iv}} &=&  \frac{1}{\sqrt{(1-\langle n_{\uparrow}\rangle)(1-\langle n_{\downarrow}\rangle)}}
 \end{eqnarray}
After making the replacements $H_{J,\text{i}} \to g_{J,\text{i}}H_{J,\text{i}} $ for each of the four terms we can rewrite the spin-contribution as $\sum_{i,\delta} \left(\tilde{J}^g_{\delta}\right)^{\mu\nu} \Sv_{i}^\mu\Sv_{i+\delta}^\nu$.  In the first two diagonal entries the $t^2$ and $B^2$ terms are renormalized by $g_{J,\text{i}}$. The Dzjaloshinskii-Moryia term is renormalized by $g_{J,\text{iv}}$ and the diagonal $A^2$ term by $g_{J,\text{iii}}$. Finally, the $(z,z)$ term is renormalized by $g_{J,\text{ii}}$. Note that although formally we have obtained $g_{J,\text{ii}}=1$ we have followed [\onlinecite{Ko}] and sent $g_{J,\text{ii}}\to g_{J,\text{i}}$ in our calculations to restore rotational symmetry that is broken by our Gutzwiller approximation. This replacement is not motivated by the specific model we have taken (where SU(2) is already broken), but instead is made to ensure that our Gutzwiller approximation hasn't broken any additional symmetry. The role of the $g_{J,\text{i}}$ factors is to approximately take $P_G$ into account on statistical grounds. There is no reason why $P_G$, which does not break rotational symmetry, should effect $\Sv_{i}^z$ and $\Sv^{x}_i$ (or $\Sv_i^y$ ) differently. The replacement $g_{J,\text{ii}}\to g_{J,\text{i}}$ is made in order to respect this fact. Note that the roman numeral subscripts on the renormalization factors {\em should not} be thought of as renormalizing specific couplings $J_1$, $J_2$, {\it etc.} nor with the (italic) site index $i$; rather they correspond to the four physical processes that can result when $\sum_{i,\delta} \tilde{J}_{\delta}^{\mu\nu} \Sv_{i}^\mu\Sv_{i+\delta}^\nu$ acts on a real space configuration of electrons. For example, $g_{J,i}$ renormalizes the spin-flip processes that occur as a result of the $J_1-J_4$ term in entries $(x,x)$ and $(y,y)$ above\cite{farrellthesis}.

The Gutzwiller renormalized spin interaction matrix reads:
  \begin{eqnarray}
 \tilde{J}^{g}_{\delta}= \left(\begin{matrix}
      \mathcal{J}_2+\mathcal{J}_3a(\delta) & 0& -\mathcal{J}_4\hat{y}\cdot \vec{\delta} \\
      0 & \mathcal{J}_2-\mathcal{J}_3a(\delta) &\mathcal{J}_4\hat{x}\cdot \vec{\delta} \\
      \mathcal{J}_4\hat{y}\cdot \vec{\delta}&-\mathcal{J}_4\hat{x}\cdot \vec{\delta}&\mathcal{J}_1\\
         \end{matrix}\right)
\end{eqnarray}

where $\mathcal{J}_1= \frac{g_{J,\text{i}}(J_1-J_2+J_4)}{2}$, $\mathcal{J}_2 = \frac{g_{J,\text{i}} (J_1-J_4)}{2}$,
$\mathcal{J}_3=\frac{g_{J,\text{iii}}J_2}{2}$ and finally $\mathcal{J}_4=\frac{g_{J,\text{iv}}J_3}{2}$. With this renormalized exchange matrix we have the following fully projected Gutzwiller Hamiltonian
\beq
H_{GA} =\sum_{\alpha, \alpha'}\sum_{ i,\delta} c^{\dagger}_{i,\alpha} \hat{t}^{g}_{\alpha,\alpha'}( \delta)c_{i+\delta,\alpha'} +  \sum_{i,\delta} (\tilde{J}_{\delta}^g)^{\mu\nu} \Sv_{i}^\mu\Sv_{i+\delta}^\nu
\eeq
we now turn our focus to a mean field study of this Hamiltonian.

\subsection{Mean-field decoupling of the projected Hamiltonian}
We are now ready to treat the approximately projected Hamiltonian in mean field.  In order to do so we define an auxiliary Hamiltonian from which the variational wavefunction $|\Psi_{BCS}\rangle$ can be found.
\begin{eqnarray}\label{Haux}
H_{Aux}&=& \sum_{\kv,\sigma} \xi_{\kv,  \alpha} c^\dagger_{\kv, \alpha} c_{\kv, \alpha} +\sum_{\kv, \alpha}
\alpha_{\kv,  \alpha} c^\dagger_{\kv, \alpha} c_{\kv,\bar{ \alpha}} \\ \nonumber  &-& \frac{1}{2}\sum_{\kv, \alpha,
 \alpha'} \left(\Delta_{\kv, \alpha,  \alpha'} c^\dagger_{\kv,  \alpha} c^\dagger_{-\kv, \alpha'}
+\text{h.c.}\right)
\end{eqnarray}
where $\xi_{\kv,  \alpha} $, $\alpha_{\kv, \alpha} $ and $ \Delta_{\kv,  \alpha, \alpha'}$ are all functions to be determined.
Note that we must allow for four channels of pairing $\Delta_{\alpha,\alpha'}$ (including triplet pairing as well) due to the spin asymmetry which is built into the model.

We perform a mean field decoupling of the Gutzwiller projected Hamiltonian $H_{GA}$ and compare the result to Equation (\ref{Haux}). In doing so we
find that the free parameters built into this auxiliary model have the following forms
\begin{eqnarray} \label{Scparam}
\xi_{\kv,\alpha} &=& \left(1-\sigma^{z}_{\alpha,\alpha}\frac{B}{t}\right)g_t(\alpha) g_{NN}\epsilon_{\kv}-\mu \\ \nonumber &+&
\hat{\xi}_{x, \alpha} \cos{k_x} + \hat{\xi}_{ y,\alpha}\cos{k_y}+ \sigma^{z}_{\alpha,\alpha}(M-4B+M_{\text{eff}}) \\ \nonumber
\alpha_{\kv,\alpha} &=& g_A g_{NN}A(\sin{k_x}-i\alpha \sin{k_y})\\ \nonumber & +&\hat{A}_x\sin{k_x}-\hat{A}_yi\sigma^{z}_{\alpha,\alpha} \sin{k_y}
\\ \nonumber
\Delta_{\kv,\uparrow,\downarrow} &=&-\Delta_{\kv,\downarrow, \uparrow}= \hat{\Delta}_x \cos(k_x)+\hat{\Delta}_y
\cos(k_y) \\ \nonumber
\Delta_{\kv,\uparrow,\uparrow} &=&-\hat{\Delta}^{\uparrow, \uparrow}_x \sin(k_x)+i\hat{\Delta}^{\uparrow,
\uparrow}_y \sin(k_y) \\ \nonumber
\Delta_{\kv,\downarrow,\downarrow} &=&\hat{\Delta}^{\downarrow, \downarrow}_x
\sin(k_x)+i\hat{\Delta}^{\downarrow, \downarrow}_y \sin(k_y) \\ \nonumber
\end{eqnarray}
where $\epsilon_\kv = -2t(\cos{k_x}+\cos{k_y})$, we have defined the effective field $M_{\text{eff}} = 2\mathcal{J}_1(\langle n_{\uparrow} \rangle -\langle
n_{\downarrow} \rangle)$ and we have defined the following parameters
\begin{eqnarray}\label{Selfconsist}
\hat{\xi}_{i,\alpha} &=& -2\mathcal{J}_2  \tilde{\xi}_{i,\bar{\alpha}}- {\mathcal{J}_1}
\tilde{\xi}_{i,\alpha}+{\mathcal{J}_4}( \tilde{A}_i+ \tilde{A}_i^*) \\ \nonumber
\hat{A}_i &=& -{2\mathcal{J}_3} \tilde{A}_{i}+{\mathcal{J}_1} \tilde{A}_{i}^*+{\mathcal{J}_4}(
\tilde{\xi}_{i,\uparrow}+ \tilde{\xi}_{i,\downarrow})  \\ \nonumber
\hat{\Delta}_i
&=&{2\mathcal{J}_2}\tilde{\Delta}_i+{\mathcal{J}_1}\tilde{\Delta}_i-{\mathcal{J}_4}\left(\tilde{\Delta}_i^{\uparrow,\uparrow}+\tilde{\Delta}_i^{\downarrow,\downarrow}\right)
\\ \nonumber
  \hat{\Delta}_i^{\alpha,\alpha} &=&{2\mathcal{J}_3}
  \tilde{\Delta}_i^{\bar{\alpha},\bar{\alpha}}-{\mathcal{J}_1} \tilde{\Delta}_i^{\alpha,\alpha}
  -{2\mathcal{J}_4} \tilde{\Delta}_i
\end{eqnarray}
In the above there are twelve free parameters that must be fixed through ensuring that the theory is
self-consistent.  These self-consistency conditions are as follows
\begin{eqnarray}\label{SCeqn}
  \tilde{\xi}_{j, \alpha}= \frac{1}{N}\sum_{\kv} \cos{k_j} \langle c_{\kv,\alpha}^\dagger c_{\kv,\alpha}\rangle
  \\ \nonumber
   \tilde{A}_j= \frac{1}{N}\sum_{\kv} q_{j} \sin{k_j} \langle c_{\kv,\uparrow}^\dagger c_{\kv,
   \downarrow}\rangle \\ \nonumber
  \tilde{\Delta}_j=  \frac{1}{N}\sum_{\kv} \cos{k_j} \langle c_{\kv,\uparrow}^\dagger c^\dagger_{-\kv,
  \downarrow}\rangle^* \\ \nonumber
    \tilde{\Delta}^{\alpha,\alpha}_j= \frac{1}{N} \sum_{\kv} \tilde{q}_{j,\alpha}\sin{k_j} \langle
    c_{\kv,\alpha}^\dagger c^\dagger_{-\kv, \alpha}\rangle^* \\ \nonumber
\end{eqnarray}
where to make the equations compact we have used $q_x=1$ and $q_y=-i$ and $\tilde{q}_{y, \alpha}=-i$ while
$\tilde{q}_{x,\alpha} = -\sigma^{z}_{\alpha,\alpha}$. These twelve equations must be solved simultaneously with the particle number
constraints $\langle n_{\alpha}\rangle = \frac{1}{N} \sum_{\kv} \langle  c_{\kv,\alpha}^\dagger
c_{\kv,\alpha}\rangle $ and $x=1-\langle n_{\uparrow}\rangle - \langle n_{\downarrow}\rangle$. This amounts to a
solving a fifteen by fifteen system of non-linear equations. For clarity let us explicitly mention the hierarchy
of notation in moving from  Eq.~(\ref{Scparam}) to Eq.~(\ref{Selfconsist}) and finally to Eq.~(\ref{SCeqn}). We have used regular
symbols, {\it e.g.} $\xi_{\kv,\alpha}$, to denote the $\kv$ dependent values in
the auxiliary Hamiltonian. The parameters with the hat symbol, {\it e.g.} $\hat{\xi}_{i,\alpha}$, are constants
that enter the definition of $\kv$ dependent functions used to define the auxiliary Hamiltonian. Finally,
we have used the tilde symbol, {\it e.g.} $\tilde{\xi}_{i,\alpha}$, for self-consistent parameters that are
heretofore unknown until we solve the equations in  Eq.~(\ref{SCeqn}).

Note that triplet superconductivity arises from the self-consistency equations whenever spin-orbit or Zeeman terms are present.
With finite spin-orbit coupling the conditions in Eq.~(\ref{SCeqn}) cannot simply be solved by setting the triplet superconductivity term $\tilde{\Delta}_j^{\alpha,\alpha}$ to zero. This stems from the
non-trivial coupling between $\hat{\xi}_{i,\sigma}$ and $\tilde{A}_i$,   $\hat{A}_{i}$ and
$\tilde{\xi}_{i,\alpha}$, $\hat{\Delta}_{i}$ and $\tilde{\Delta}^{\alpha,\alpha}_i$ and
$\hat{\Delta}^{\alpha,\alpha}_i$  and $\tilde{\Delta}_{i}$ that is in place as a results of the coupling
$\mathcal{J}_4$. This is seen by inspecting the last term in each of the definitions in Equation
(\ref{Selfconsist}). To see why this coupling refuses solutions where some of the parameters are identically
zero consider looking for a solution where $\tilde{\Delta}_i^{\alpha,\alpha}=0$. The last two equations in Eq.~(\ref{Selfconsist}) then become
\begin{eqnarray}
\hat{\Delta}_i &=&-{2\mathcal{J}_2}\tilde{\Delta}_i-{\mathcal{J}_1}\tilde{\Delta}_i\\ \nonumber
  \hat{\Delta}_i^{\alpha,\alpha} &=&{2\mathcal{J}_4} \tilde{\Delta}_i
\end{eqnarray}
and the relevant equations we must solve are
\begin{eqnarray}\label{nonzero}
  \tilde{\Delta}_j=  \frac{1}{N}\sum_{\kv} \cos{k_j} \langle c_{\kv,\uparrow}^\dagger c^\dagger_{-\kv,
  \downarrow}\rangle^* \\ \nonumber
0= \frac{1}{N} \sum_{\kv} \tilde{q}_{j, \sigma}\sin{k_j} \langle c_{\kv,\alpha}^\dagger c^\dagger_{-\kv,
\alpha}\rangle^* \\ \nonumber
\end{eqnarray}
It is  impractical to find analytic expressions for averages such as $\langle c_{\kv,\alpha}^\dagger
c^\dagger_{-\kv, \alpha}\rangle$ (they depend on eigenvectors of a complicated four by four matrix) however we
can find these numerically. Looking at numerics and using some intuition we argue that very roughly $\langle
c_{\kv,\alpha}^\dagger c^\dagger_{-\kv, \alpha}\rangle \propto \hat{\Delta}_i^{\alpha, \alpha}$ and we have said
above that $\hat{\Delta}_i^{\alpha, \alpha}\sim \tilde{\Delta}_i$. So in general either $\tilde{\Delta}_i=0$ or
the sum on the right of the second equation in (\ref{nonzero}) is nonzero and therefore
$\tilde{\Delta}_i^{\alpha,\alpha}$ cannot be zero.

We have solved the equations developed in the theory above numerically. Where these parameters are non-zero the
general trend we see in the solutions is $|\hat{\xi}_{x,\alpha}| = |\hat{\xi}_{y,\alpha}| = \hat{\xi}_{\alpha}$,
$\hat{A}_x=\hat{A}_y = \hat{A}$, $\hat{\Delta}_x = - \hat{\Delta}_y$ and $\hat{\Delta}^{\alpha,\alpha}_x = -
\hat{\Delta}_y^{\alpha,\alpha}$. This equality of parameters with $x$ and $y$ directionality is not surprising
as our model has no preference for the $x$ or $y$ direction.

\begin{figure}[tb]
  \setlength{\unitlength}{1mm}

   \includegraphics[scale=.6]{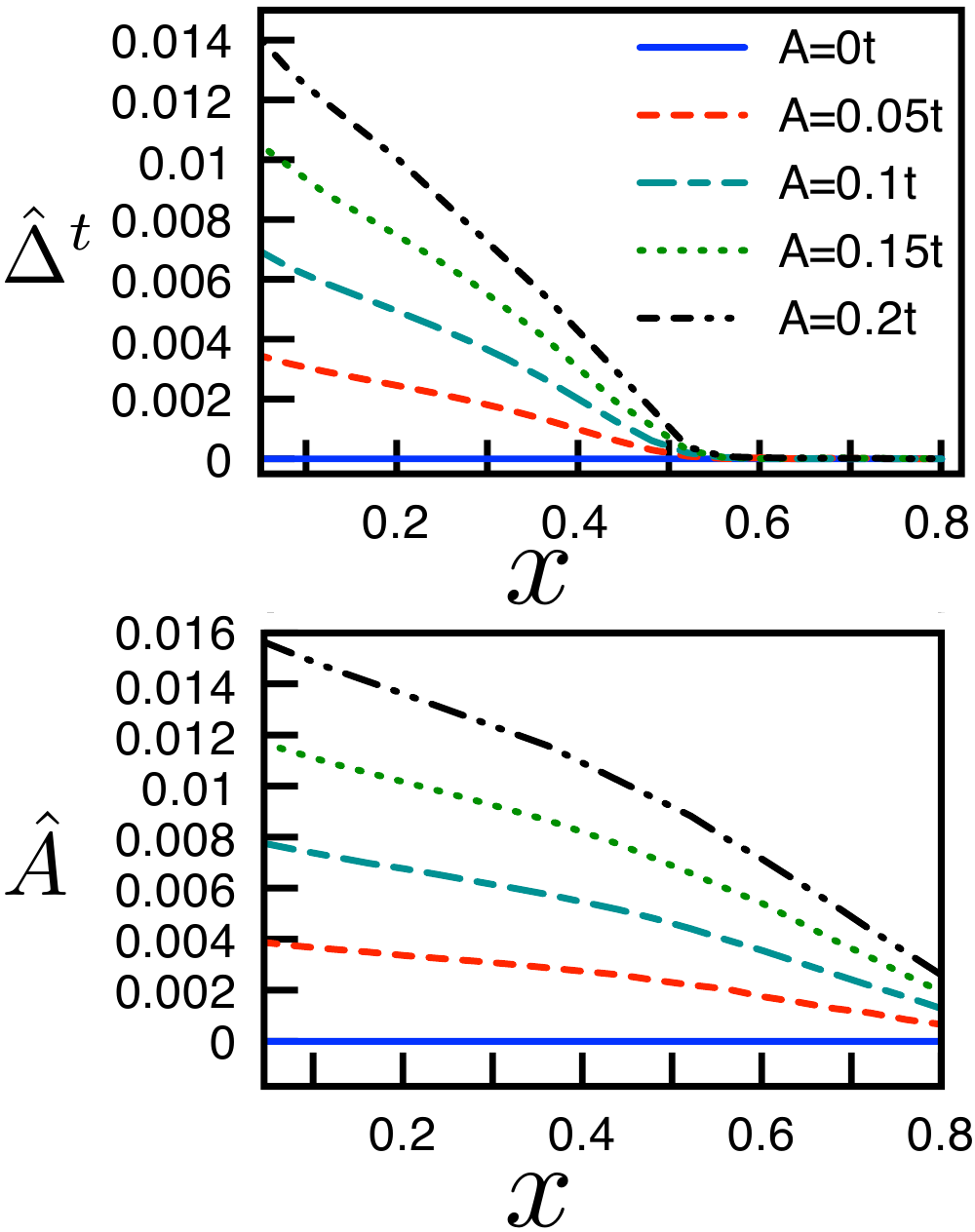}
\caption{{\small
Plot of the parameters $\hat{\Delta}^{t}$ and $\hat{A}$ for various values of the bare spin-orbit coupling
parameter $A$. We have kept $M=B=0$ in this plot and the order parameters are plotted in energies of $t$. We
have set $U=12t$ and $V=-0.05U$ in this figure.
     }
     }\label{fig:SC_f_A}
\end{figure}

We first present our results for $M,B=0$ and $A\ne0$. In this special case of the spin-orbit coupling parameters
there is not preferred spin direction and we have $\hat{\xi}_{\alpha}= \hat{\xi}$, $\langle n \rangle _{\alpha}
= (1-x)/2$ and $\hat{\Delta}_i^{\uparrow,\uparrow}= \hat{\Delta}_i^{\downarrow,\downarrow}= \hat{\Delta}^t_i$.
Further, when we solve the self-consistency conditions we find that the values of $\hat{\xi}$ and
$\hat{\Delta}$, the kinetic energy and $d$-wave superconductivity parameters, are effected very little by
letting $A$ become finite, their behaviour remains as described in the previous section and plots of their behaviour is presented in Appendix~\ref{ap:NoSOC-BCS} in Fig. \ref{fig:OP}. In view of this, in
Fig.~\ref{fig:SC_f_A} we have plotted the doping dependance of $\hat{\Delta}^{t}$ and $\hat{A}$ for various
values of $A$.

In Fig.~\ref{fig:SC_f_A} for $A=0$, $\hat{\Delta}^{t}$ and $\hat{A}$ are zero for all
values of doping. As we allow $A$ to become finite both develop behaviour that is monotonically decreasing in
the doping $x$. Like $\hat{\xi}$, $\hat{A}$ is finite over the whole range of $x$ values shown while $\hat{\Delta}$ aand $\hat{\Delta}^{t}$ vanish together at some critical $x_c$ which is unaffected by $A$. Both  $\hat{\Delta}^{t}$ and $\hat{A}$ increase in magnitude with
$A$ but their general shape is unaffected. We also note that the values of these parameters is an order of
magnitude less than their counterparts   $\hat{\xi}$ and $\hat{\Delta}$.

\subsection{Self-Consistent Parameters}
We note that with a finite value of $M$ the system will experience some spin-polarization and there is no reason to expect $\langle n_{\uparrow} \rangle = \langle n_{\downarrow} \rangle $. In this case we must solve all $15$ self-consistent equations derived in the previous section of this paper. Proceeding to do so we find that the following results hold $\hat{\xi}_{x,\sigma} = \hat{\xi}_{y,\sigma}= \hat{\xi}_{\sigma}$, $\hat{A}_x=\hat{A}_y = \hat{A}$, $|\hat{\Delta}_x| = | \hat{\Delta}_y|=\Delta_{s}$ and $|\hat{\Delta}^{\sigma,\sigma}_x| = | \hat{\Delta}_y^{\sigma,\sigma}|= \Delta_{t}^{\sigma}$.

\begin{figure*}[tb]
  \setlength{\unitlength}{1mm}

   \includegraphics[scale=.45]{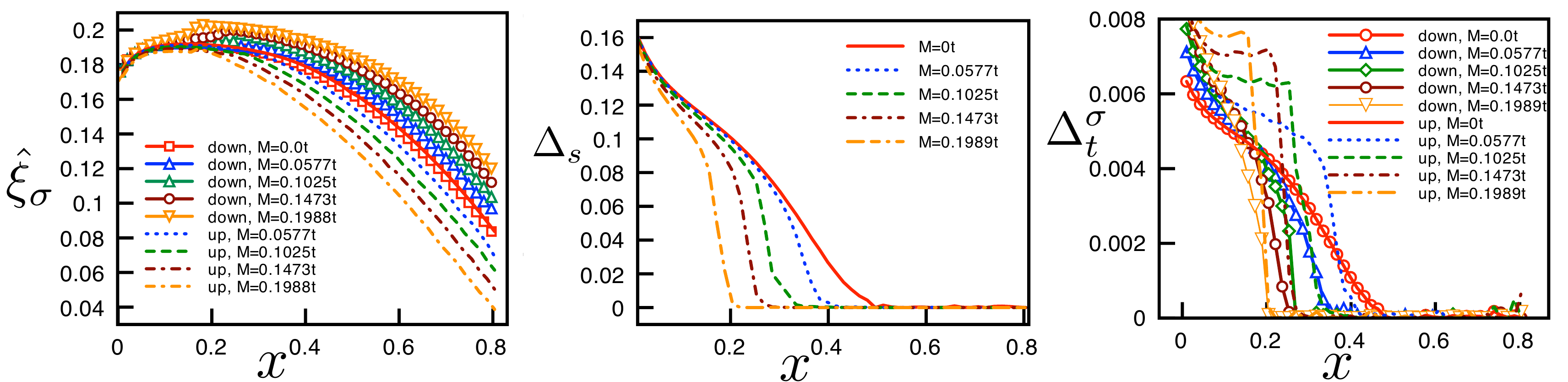}
\caption{{\small
Self-consistent parameters for a system at finite Zeeman field. In calculating the data above we have set $B=0$, $U=12t$, $V=-0.05U$ and $A=0.1t$ while allowing $M$ to vary. The values of $M$ used are labelled in the legend of the figure. Where the legend indicates a direction (up or down) this refers to the value for $\sigma$ in either $\hat{\xi}_\sigma$ or $\Delta_{t}^{\sigma}$. Note that the up and down data at $M=0$ are indistinguishable.
     }
     }\label{fig:finite_M}
\end{figure*}

To explore the implications of a non-zero $M$ we have obtained data for $U=12t$, $V=-0.2U$ and $A=0.1t$ as a function of both $M$ and $x$. Our results are shown in Figure \ref{fig:finite_M}. In Fig. \ref{fig:finite_M}(a) we have plotted both $\hat{\xi}_{\uparrow}$ and  $\hat{\xi}_{\downarrow}$. Starting at $M=0$ we have $\hat{\xi}_{\uparrow}=\hat{\xi}_{\downarrow}$ and the hopping amplitude of both spin-up and spin-down electrons is renormalized in the same way. As we move away from $M=0$ we see that  the renormalization changes  so that $\hat{\xi}_{\uparrow}<\hat{\xi}_{\downarrow}$ with the difference between the two renormalizations increasing with $M$. Therefore for finite $M$ the renormalized hopping amplitude for spin-up and spin-down electrons are different.

In Fig.~\ref{fig:finite_M}(b) now we see how changing $M$ changes the spin-singlet pairing parameter $\Delta_s$. For $M=0$ this parameter goes smoothly to zero for $x>x_c$. As we allow for a finite $M$ we see that the amplitude of this pairing parameter is not altered significantly for small $x$. The drastic difference comes in when we look at how and when $\Delta_s$ goes to zero at higher $M$. There the smooth transition to $\Delta_s=0$ is replaced by a sharper drop. The transition becomes sharper as $M$ increases. The other marked difference between the $M=0$ and the $M\ne0$ case is the critical filling $x_c$ at which $\Delta_s\to0$. We see in the figure that as $M$ is increased $x_c$ becomes smaller.

Fig.~\ref{fig:finite_M}(c) gives both the up-up and down-down triplet pairing parameter for various values of $M$. Again starting at the $M=0$ limit we see that both pairing amplitudes are identical over all $x$ and smoothly go to zero at some $x_c$. Once we turn the Zeeman field on a difference between the up and down pairing parameter develops. Both being identical at half-filling ({\it i.e.} $x=0$) but as the hole doping is increased the down pairing value becomes larger than the up pairing amplitude and like $\hat{\xi}_{\sigma}$ the difference between the two becomes larger for larger $M$. Further, the continuous vanishing of $\Delta_{t}^{\sigma}$ at zero $M$ disappears as we increase $M$ and both the $\Delta_{t}^{\sigma}$ discontinuously drop to zero. Like with $\Delta_{s}$ the critical value of $x$ at which this drop occurs decreases as $M$ increases.

\subsection{Topology}
The task of calculating the topological invariant in an interacting system is not at all a simple one.  There are several methods one can explore\cite{interactingTI0,interactingTI1,interactingTI2,interactingTI3,interactingTI4,interactingTI5} however, all of them are very involved numerically.  In this work we choose circumvent this difficulty by making a plausible conjecture.  We argue that the topology of our ground state, which is a projected mean field state, is the same as the topology of the mean-field state {\it before} the projection ({\it i.e.} the ground state of the Hamiltonian in Eq. (\ref{Haux})).  To see why this is plausible consider the electronic correlations in a topological system.  In the bulk we expect the quasiparticle correlations to decay exponentially with the distance due to the bulk excitation gap.  However, on the edges the topology guaranties protected edge modes.  The correlation on the edge is therefore long ranged and is characterized by a power law.  Suppose that we have a non-interacting ground state whose topological invariant is non-zero.  It's correlations are as described above.  If we now apply the Gutzwiller projection to this state, being a local operation we do not expect it to alter the electronic correlations.  We therefore conclude that its topology is unchanged by the projection.  This is a subtle issue that awaits further study.

We therefore proceed to calculate the topological invariant of the un-projected BCS state which satisfies the self-consistency equations we found above.  This is equivalent to determining the topology of the mean field Hamiltonian $H_{Aux}$.
In order to calculate the topological invariant of $H_{Aux}$ we appeal to an extremely useful method developed in Ref. [\onlinecite{Ghosh}], which we do not review here.
These authors show, based on symmetry arguments that, on a square lattice the topology is expressed by the first Chern number, $C_1$, which can be elegantly written as:
\beq
C_1 = \frac{1}{i\pi} \log\left(\frac{Q(\mathcal{H}(0, 0)) Q(\mathcal{H}(\pi, \pi))}{Q(\mathcal{H}(\pi, 0)) Q(\mathcal{H}(0, \pi))}\right)
\eeq
where the $Q$ is defined as
\beq
Q(\mathcal{H}(\kv)) = \text{sgn}\left(-\text{Pf}\left[\mathcal{H}(\kv)\Lambda\right]\right)
\eeq
where $\Lambda = \sigma_y \otimes \tau_y$ ($\sigma$ and $\tau$ are Pauli matrices acting on the spin and Nambu spaces respectively) is a symmetry of our Hamiltonian and $\mathcal{H}(\kv)$ is the Bogoliubov de Gennes (BdG) Hamiltonian obtained from writing $H_{Aux}$ in the basis $(c_{\kv,\uparrow}, c_{\kv,\downarrow}, c^\dagger_{-\kv,\downarrow}, -c^\dagger_{-\kv,\uparrow})^T$ .
Taking the definition of $\log(z)$ such that $\log(1)=0$ and $\log(-1)=i\pi$ and noting that the $Q$-invariants only ever have values of one or minus one we see that $C_1$ takes only two distinct values: $C_1=0$ for trivial topology and $C_1=1$ for non-trivial topology.

Evaluating the 4 relevant $Q$ invariants for our system we find that the topological invariant $C_1$ is given by
\beq\label{chern}
C_1 = \frac{1}{i\pi} \log\left(\text{sgn}\left[(\eta_\uparrow^2-(V_z-\mu)^2)(\eta_\downarrow^2-(V_z+\mu)^2)   \right]\right)
\eeq
where we have defined $\eta_{\sigma} = \hat{\xi}_{x,\sigma}+\hat{\xi}_{y,\sigma} -4tg_t(\sigma)g_{NN}-4\sigma Bg_t(\sigma)g_{NN}$ and the effective Zeeman term $V_z = M+M_\text{eff}-4B$. We now notice that given the above definition for $C_1$ we can easily ``sort" our parameter space into 4 distinct categories depending on the values of $ {\eta}_{\sigma}$, $\mu$ and $V_z$. The first two categories are $|\eta_{\downarrow}| > |V_z+\mu|$ and  $|\eta_{\uparrow}| > |V_z-\mu|$ (category 1) or $|\eta_{\downarrow}| < |V_z+\mu|$ and  $|\eta_{\uparrow}| < |V_z-\mu|$. If our parameters lay in either of these categories then $C_1=0$ and the system has trivial topology. Meanwhile if $|\eta_{\downarrow}| < |V_z+\mu|$ and  $|\eta_{\uparrow}| > |V_z-\mu|$ (category 3) or $|\eta_{\downarrow}| > |V_z+\mu|$ and  $|\eta_{\uparrow}| < |V_z-\mu|$ (category 4) then $C_1=1$ and the system has non-trivial topology.

Notice that although the bare parameters $A, U$ and $V$ do not appear explicitly in the definition of $C_1$ in Eq.~(\ref{chern}) the parameters $\mu$, $\xi_{i,\sigma}$ and $M_\text{eff}$ have been determined self-consistently and therefore depend implicitly on $A, U$ and $V$. Further, we see that $C_1$ depends on the doping $x$ both explicitly through the renormalization parameters $g_t(\sigma)$ and $g_{NN}$ and implicitly through the self-consistent study completed to find $\mu$, $\hat{\xi}_{i,\sigma}$ and $M_\text{eff}$. In this way $C_1$ depends on all of the parameters built into the model in a non-trivial way.

It is clear from the above definition of $C_1$ that for a non-trivial phase either $B$ or $M$ must be non-zero. To see this suppose we have set $B=M=0$, then there is no spin polarization and therefore $M_\text{eff}=0$ and $V_z=0$. Further for $M,B=0$ we have $\xi_{i,\sigma} = \hat{\xi}$ and therefore $\eta_\uparrow=\eta_{\downarrow}=\eta$. Thus the argument inside the log becomes $(\eta^2-\mu^2)^2$ which is positive definite so that $C_1=0$ regardless of $A/t$, $V_0/t$, $U/t$ or $x$.

\begin{figure*}[tb]
  \setlength{\unitlength}{1mm}

   \includegraphics[scale=.45]{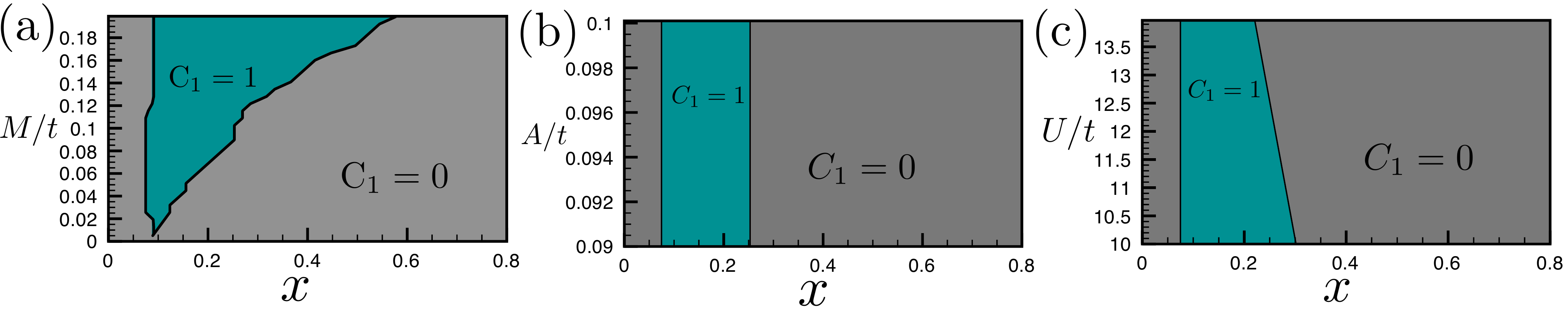}
\caption{{\small
Topological phase diagram for several different parameters cuts. In (a) we have set $U=12t$, $B=0$, $V=-0.05U$ and $A=0.1t$ and varied $M$ and $x$, in (b) we fix $U=12t$, $B=0$, $V=-0.05U$ and $M=0.1t$ while varying $A$ and $x$ and in (c) $M=.1t$, $B=0$, $V=-0.05U$ and $A=0.1t$ while varying $U$ and $x$.
     }
     }\label{fig:Top_PD}
\end{figure*}

Our results for the topological invariant as a function of the model's bare parameters are shown in Fig.~\ref{fig:Top_PD}.

 \section{Conclusions}

We have studied the strong interaction limit of the extended Hubbard model with spin-orbit coupling.
Performing a strong coupling expansion we have found an effective Hamiltonian for this model valid in the limit
$U, V \gg t,A,B$. We have shown that by projecting this model unto a subspace of half-filled many-body
states we obtain a spin model that is a generalization of the Heisenberg model with Dzjaloshinskii-Moriya and compass anisotropy originating from the spin orbit coupling terms in the extended Hubbard model.  This model exhibits an interesting spin texture which will be explored elswhere\cite{SpinHamiltonian}. We then set out to study this effective Hamiltonian {\it via} mean field theory. When the filling is lower than half the system may become superconducting and we study the various superconducting phases in the framework of the Gutwiller approximation.
When we study the system in the presence of spin-orbit and Zeeman terms we find a variety of superconducting order parameters both in the singlet and in the triplet channels.  The topology of the superconducting state is found to be non trivial at some finite doping and a phase diagram is given.

\section{Acknowledgements}

 We would like to thank E. Altman, A. Paramekanti and N. Trivedi for useful discussions. Financial support for this work was provided by the NSERC and FQRNT (TPB) the Vanier Canada Graduate Scholarship (AF), the Schulich Graduate Fellowship (AF), the Walter C. Sumner Memorial Fellowship (AF) and the Tomlinson Master's Fellowship (AF). The majority of the numerical calculations were performed on CLUMEQ supercomputing resources.  TPB wishes to acknowledge the hospitality of the Aspen center for Physics where some of this work has been done.

\bibliographystyle{apsrev}
\bibliography{topoSC}

\appendix
\section{Strong Coupling Expansion}\label{ap:expansion}
\subsection{Interaction as a Constant of Motion and Channel Decomposition}

Here we will give the full details of the strong coupling expansion discussed in Section \ref{sec:SCE}. For an in depth discussion of the details in this appendix and Appendix \ref{ap:NoSOC-BCS} please see Ref. [\onlinecite{farrellthesis}].  We begin with the full (untransformed) Hamiltonian which reads
\begin{equation}
H = H_1+H_0+H_Z.
\end{equation}
where $H_0=H_{int}=H_U+H_V$, $H_U=U\sum_{i} n_{i,\uparrow}n_{i,\downarrow}$, $H_V = V\sum_{\langle i,j\rangle} n_in_j$ and $H_1=T+H_{SO}$. As advertised in the text we wish to perform a unitary transformation on the above Hamiltonian in such a way as to block diagonalize $H$. An alternative, but entirely equivalent, approach to this is to insist that $H_0$ is a constant of motion in the transformed Hamiltonian. To this end we write the following unitary transformation of $H$
\begin{equation}
\tilde {H} = e^{iS} (H_1+H_0+H_Z) e^{-iS} = H_0+\tilde{H}_1+H_Z,
\end{equation}
where $H_Z$ should not change under the transformation $S$ since it does not change $H_0$. In order for $H_0$ to be a constant of motion we require that
\begin{equation}\label{constant}
[H_0, \tilde {H}] = [H_0, \tilde{H}_1]=0.
\end{equation}
To satisfy this condition we expand both $\tilde{H}_1$ and the transformation $S$ in a power series of $1/U$ as follows
\begin{equation}
S=-i\sum_{n=1}^\infty \frac{S_n}{U^n}, \ \ \ \ \ \ \ \tilde{H}_1 = \sum_{n=1}^{\infty} \frac{\tilde{H}_{1,n}}{U^{n-1}}
\end{equation}
and then insist that Eq.~(\ref{constant}) is satisfied to a given order in $1/U$, {\it i.e.} $[H_0, \tilde{H}_{1,n}]=0$
for some $n\ge1$.

To proceed with solving the condition in Eq.~(\ref{constant}) it is useful to decompose $H_1$ into channels that
change $H_0$ by a given amount. We recall that the first term in $H_0$, $H_U$, counts the number of doubly occupied sites.
We sort the hopping processes into channels which change the number of doubly occupied sites by $m$ (changing the on site interaction energy by $mU$). In a single band model we have only three possible values $m=-1,0,1$. Using the electron and hole occupancy operators, $h_{i,\alpha}=1-n_{i,\alpha}$ we find:
 \begin{eqnarray}
  \ {T}_{-1} &=& \sum_{i,\sigma,\delta, \sigma'} h_{i,\bar{\sigma}} c^\dagger_{i,\sigma}
  \hat{t}_{\sigma,\sigma'}(\vec{\delta})c_{i+\delta,\sigma'} n_{i+\delta,\bar{\sigma}'} \\ \nonumber
 \ {T}_{1} &=& \sum_{i,\sigma,\delta, \sigma'} n_{i,\bar{\sigma}} c^\dagger_{i,\sigma}
 \hat{t}_{\sigma,\sigma'}(\vec{\delta})c_{i+\delta,\sigma'} h_{i+\delta,\bar{\sigma}'} \\ \nonumber
     \ {T}_{0} &=& \sum_{i,\sigma,\delta, \sigma'} (n_{i,\bar{\sigma}} c^\dagger_{i,\sigma}
     \hat{t}_{\sigma,\sigma'}(\vec{\delta})c_{i+\delta,\sigma'}  n_{i+\delta,\bar{\sigma}'} \\ \nonumber &&+
     h_{i,\bar{\sigma}} c^\dagger_{i,\sigma} \hat{t}_{\sigma,\sigma'}(\vec{\delta})c_{i+\delta,\sigma'}
     h_{i+\delta,\bar{\sigma}'})
 \end{eqnarray}
and this definition also leads to $[T_m, H_U]=mUT_m$.

In a similar way we should further classify the terms in $H_1$ according to their effect on the number of nearest neighbor pairs.
This is done with the help of the the nearest neighbor projection operator
 \begin{eqnarray}
 {O}_i(\tilde{n}_{x,\uparrow}, \tilde{n}_{x,\downarrow},...\tilde{n}_{-y,\downarrow}) &=& \prod_{\delta,\alpha}
 (\tilde{n}_{\delta,\alpha}n_{i+\delta, \alpha}+(1-\tilde{n}_{\delta,\alpha})h_{i+\delta,\alpha}) \nonumber \\
 &\equiv& {O}_i[n]
 \end{eqnarray}
which projects out all possible orientations of the nearest neighbors of atom site $i$ except the one labeled
by $n$ where the arguments $n_{\delta,\alpha}$ take a value of 1 (a value of 0) for site $i+\delta$ begin
occupied (unoccupied). Inserting the identity $1=\prod_{\delta, \alpha}
(n_{i+\delta,\alpha}+h_{i+\delta,\alpha})$ on either side of the summand in the channels ${T}_m$ we obtain the full decomposition operator
\begin{equation}
\ {T}_{m,N_2,N_1} =\small{ \sum_{i,\delta,\sigma, \sigma'} \sum_{\Sigma[n_1]=N_1} \sum_{\Sigma[n_2]=N_2} O_i[n_2]
({T}_m)_{i,\delta,\sigma, \sigma'}O_{i+\delta}[n_1] }
\end{equation}
where ${T}_{m,N_2,N_1}$ changes the number of doubly occupied sites by $m$ while and the number of nearest neighbors from $N_1$ to $N_2$. $\Sigma[n] \equiv \sum_{\delta, \alpha} n_{\delta,\alpha}$.

A tedious but straightforward calculation shows that $ [\ {T}_{m,N_2,N_1}, H_U+H_V] = (mU+(N_2-N_1)V)\ {T}_{m,N_2,N_1}$.

\subsection{The Transformed Hamiltonian}

To obtain an effective Hamiltonian we must solve for the transformation which satisfies $[H_0, H'_{1,n}]=0$ to a
given order $n$.  We then use this transformation to find the properly expanded
Hamiltonian. To do this we first note that for any operator $X$
\begin{equation}
e^{iS} X e^{-iS}= X+[iS,X]+\frac{1}{2} [iS,[iS,X]]+...
\end{equation}
and so we have
\begin{equation}
\tilde {H} = H_0 +(H_1+[iS, H_0]) +([iS,H_1]+\frac{1}{2}[iS, [iS, H_0]])+...
\end{equation}
Inserting the expansions for $S$ into the above and redefining $H_0=U\tilde {H}_0$ we find
\begin{eqnarray}\label{conds}
&&\tilde{H}'_{1,1} = H_1 + [S_1, \tilde {H}_0]  \\ \nonumber &&\tilde{H}'_{2,1} = [S_1, {H}_1] + [S_2, \tilde {H}_0]
+\frac{1}{2}[S_1,[S_1, \ {H}_0]]
\end{eqnarray}
The first condition we must satisfy is then
\begin{equation}
0 = [{H}_0, H_1 + [S_1, \tilde {H}_0]]
\end{equation}
Inserting the channel decomposition $H_1 = \sum_{m,N_2,N_1}{T}_{m, N_2,N_1}$ and using the fact that $[\tilde
{H}_0, H_1] =  \frac{1}{U}\sum_{m,N_2,N_1}(mU+(N_2-N_1)V)\ {T}_{m, N_2,N_1}$ the above is then solved by
\begin{equation}
S_1 = \sum'_{m,M,N} \frac{U\ {T}_{m, N_2,N_1}}{mU-(N_2-N_1)V}
\end{equation}
where the primed sum excludes terms for which $mU-(N_2-N_1)V=0$. Using this expression in the first order
Hamiltonian we have
\begin{equation}
H'_{1,1} = \sum_{m,N_2,N_1}\ {T}_{m, N_2,N_1} -  \sum'_{m,N_2,N_1}\ {T}_{m, N_2,N_1} = \sum_N \ {T}_{0,N,N}
\end{equation}
Let us pause and examine this result in view of our goals at the onset of this discussion. This tells us to
order $\mathcal{O}(1)$ the piece of the full Hamiltonian that does not change $H_U+H_V$ involves the sum of all
channels with hops that do not change the number of sites with double occupancy and begin and end with $M$ occupied
nearest neighbor bonds. We see that none of these operations change $H_U$ or $H_V$ and we could have argued the
result above on heuristic grounds.

Next we must satisfy the condition $[H_0, H'_{2,1}]=0$ by properly choosing $S_2$. The mathematical expression
for $S_2$ is very complicated and not necessary for our discussion. We therefore simply present the result which is the second order contribution to $H_1'$
\begin{equation}
H'_{1,2} = \sum_{m,M_1, N_1, M_2}' \frac{U [\ {T}_{m,M_1, N_1}, \ {T}_{-m, M_2,
M_1+M_2-N_1}]}{2(mU+(M_1-N_1)V)}
\end{equation}
where the sum is over all indices except ones for which $mU+(M_1-N_1)V=0$. Again we can inspect this term
and see that it is precisely what we would have guessed from the onset. First, this term is quadratic in the
$T_{m,N_2,N_1}$ operators and so it represents a second order term in our expansion. Second, the specific
combination of the $T_{m,N_2,N_1}$ operators does not change the number of doubly occupied sites nor the number
of occupied nearest neighbor bounds; ${T}_{-m, M_2, M_1+M_2-N_1}$ decreases the number of double occupied sites
by $m$ and changes the number of occupied nearest neighbor bounds by $N_1-M_1$ and then ${T}_{m,M_1, N_1}$
increases the number of doubly occupied sites by $m$ and changes the number of occupied nearest neighbors by
$M_1-N_1$.

Using the two results above we can now write the effective Hamiltonian that has been the focus of this paper. Valid to order $\mathcal{O}(1/U^2)$ this effective Hamiltonian is
\begin{eqnarray}
\tilde {H} &=& H_U+H_V + H_Z + \sum_{M} \ {T}_{0,M,M} \\ \nonumber &+& \sum_{m,M_1,N_1,M_2}' \frac{ {T}_{m,M_1,
N_1}  {T}_{-m, M_2, M_1+M_2-N_1}}{(mU+(M_1-N_1)V)}+\mathcal{O}(1/U^2),
\end{eqnarray}
where we made use of index relabelling symmetry to get rid of the commutator and the factor of a half in the
second term of the expansion. In principal we could keep going to higher orders in $1/U$. The next order term
would involve cubic combinations of the $T_{m,N_2,N_1}$ operators that collectively do now change the number of
doubly occupied sites nor the number of occupied nearest neighbor bonds. We stop here for practicality as these
higher order terms become very complicated to write down and are not needed to accomplish our current goal.

\subsection{Projection to Half-Filling: An analogy to the $t-J$ model}
An expansion analogous to the one performed in the previous subsection is used to write an effective
strong coupling theory for the Hubbard model\cite{macdonald}. If one then projects this strong coupling Hubbard model to half-filling one
obtains a model that amounts to the Heisenberg model for spins on a lattice.\cite{Chao} This is typically how one shows
that the strongly interacting limit of the Hubbard model is the so-called $t-J$ model. For the sake of
comparison, we will follow a similar process in this section in order to obtain an effective model for
$\tilde{H}$ that is applicable at half filling.

Let us begin by considering some state $|\psi\rangle$ at half-filling. For strong $U$ the electrons will avoid
configurations with double occupancies and the state $|\psi\rangle$ will consist of a lattice of singly occupied
sites. Let us consider the action of the following term on the subspace of possible states $|\psi\rangle$
\begin{equation}
 \sum_{m,M_1,N_1,M_2}' \frac{ {T}_{m,M_1, N_1}  {T}_{-m, M_2, M_1+M_2-N_1}}{(mU+(M_1-N_1)V)}.
\end{equation}
To begin with $|\psi\rangle$ has only singly occupied sites and so the only possible value of $-m$ is $-m=1$ as
in this scenario we cannot maintain the number of double occupancies ($m=0$) nor can we decrease the number of
double occupancies ($m=1$). Next, all sites in $|\psi\rangle$ have $4$ occupied nearest neighbors and so
$M_1+M_2-N_1=4$. Next, after a hop has occurred the electron will be on a site with $3$ nearest neighbors and
so $M_2=3$. If we want to begin and end in a subspace of all possible half-filled states with no double
occupancies the second part of the term above, namely $ {T}_{m,M_1, N_1}$, must now destroy the doubly occupied
site ${T}_{-m, M_2, M_1+M_2-N_1}$ has created. For this to be the case we must have $m=-1$ and $N_1=3$ which
leads to $M_1=4$. Plugging in all of these results we get the following
\begin{equation}
 \sum_{m,M_1,N_1,M_2}' \frac{ {T}_{m,M_1, N_1}  {T}_{-m, M_2, M_1+M_2-N_1}}{(mU+(M_1-N_1)V)}\to -\frac{
 {T}_{-1,4, 3}  {T}_{1, 3,4}}{U-V}
\end{equation}
Taking all of our projections into account and inserting the definitions of $T_{m,N_2,N_1}$ into the above we
then find
\begin{equation}\label{spin2}
 \frac{ {T}_{-1,4, 3}  {T}_{1, 3,4}}{U-V} =\small{\sum_{i,\delta,\beta, \beta',
 \alpha,\alpha'}\frac{c^\dagger_{i+\delta,\alpha} \hat{t}_{\alpha,\alpha'}(-\vec{\delta})c_{i,\alpha'}
 c^\dagger_{i,\sigma} \hat{t}_{\beta,\beta'}(\vec{\delta})c_{i+\delta,\sigma'}}{U-V}  }
\end{equation}
where we have dropped all projection operators because we have already taken them into account. The spin sums in
the above term can be eliminated if we write things in terms of the lattice spin operator given by the
following
\begin{equation}\label{sitespins}
(\Sv)_i = \frac{1}{2} \sum_{\alpha,\beta} c^\dagger_{i,\alpha} \vec{\sigma}_{\alpha,\beta} c_{i,\beta}
\end{equation}
Performing all of these spin sums is tedious but straightforward. We will describe the process for the specific term $\beta=\bar{\beta}'$ with $ \alpha=\bar{\alpha}'$ sum
and then simply quote the result for all of the other sums.

Setting $\beta=\bar{\beta}'$ and $ \alpha=\bar{\alpha}'$ in the summand of Eq.~(\ref{spin2})  gives us the
following contribution
\begin{equation}
 \sum_{i,\delta,\beta, \alpha,}\frac{A^2(\delta_x+\sigma^{y}_{\beta,\bar{\beta}}\delta_{y})
 (\delta_x+\sigma^{y}_{\alpha,\bar{\alpha}}\delta_{y})}{4(U-V)}c^\dagger_{i+\delta,\alpha}c_{i,\bar{\alpha}}
 c^\dagger_{i,\beta}c_{i+\delta,\bar{\beta}}
\end{equation}

We first consider the case where $\vec{\delta}=\pm\hat{x}$, in this case we have, after making one commutation
\begin{equation}
 n_i-\sum_{\beta, \alpha} c^\dagger_{i+\delta,\alpha}c_{i+\delta,\bar{\beta}}
 c^\dagger_{i,\beta}c_{i,\bar{\alpha}}
\end{equation}
where we have only written the sum over spins and for brevity we have dropped the prefactor $A^2/(4(U-V))$ which
will be added back later. Writing out all four terms in the sum above, using the definition in Eq.~(\ref{sitespins}) to identify spin operators, recalling that we are interested on the action of our Hamiltonian
on the subspace of half-filled singly occupied states and doing a few manipulations and  we arrive at
 \begin{equation}
 2\left((S_y)_{i+\delta}(S_y)_{i}-(S_x)_{i+\delta}(S_x)_{i}+(S_z)_{i+\delta}(S_z)_{i}+\frac{n_in_{i+\delta}}{4}\right).
\end{equation}

Completing the calculation following the above formula we arrive at the following result
\begin{equation}
 \frac{ {T}_{-1,4, 3}  {T}_{1, 3,4}}{U-V}  =E_0 +\sum_{i,\delta} J_\delta^{\mu\nu} \Sv_{i}^\mu \Sv_{i+\delta}^\nu
\end{equation}
where we have defined the constant
\begin{equation}
E_0 =-\frac{2Nt^2}{U-V} - \frac{NA^2}{2(U-V)}-\frac{2NB^2}{U-V}
\end{equation}
and the $3\times3$ matrix
\begin{widetext}
\begin{eqnarray}
 J_{\delta}= \frac{1}{2(U-V)} \small{\small{\left(\begin{matrix}
      4t^2+{A}^2a(\delta)-4B^2 & 0& -4At\delta_y \\
      0 & 4t^2-{A}^2a(\delta)-4B^2&4At\delta_x \\
      4At\delta_y&-4At\delta_x&4t^2-A^2+4B^2\\
   \end{matrix}\right)}}
\end{eqnarray}
\end{widetext}
where $a(\delta)$ is $1$ for $\delta=\pm \hat{x}$ and $-1$ for $\delta=\pm\hat{y}$. Note that the above
result readily simplifies to the Heisenberg model for spins on a lattice in the limit $V,A,B\to0$ as $J_\delta$
becomes the identity matrix times $2t^2/U$.

To obtain our analogy to the $t-J$ mode we drop the overall constant $E_0$ from the
above result, we also drop $H_U$ from the results in $\tilde{H}$ because near half filling and at strong
coupling the number of doubly occupied sites will be zero and finally we drop $H_V$ as the number of occupied
nearest neighbours should be constant, $2N$, on a square lattice. Thus our effective model is given by
\begin{equation}\label{Atjd}
H_{t,J_\delta} = H_Z + \sum_{N} T_{0,N,N} +\sum_{i,\delta} J_\delta^{\mu\nu} \Sv_{i}^\mu \Sv_{i+\delta}^\nu.
\end{equation}
Compared to the $t-J$ model we have several obvious differences. First, with spin-orbit coupling the exchange
constant $J$ has now become a matrix. Second, instead of hopping that is restricted to not change the number of
doubly occupied sites in the $t$ part of the $t-J$ model, our term $\sum_{M} T_{0,M,M}$ not only restricts
changes in the number of doubly occupancies but also the number of occupied nearest neighbors. Lastly, because
of the Zeeman/mass term in the original model we have the magnetic like contribution $H_Z$.

\section{Mean field BCS treatment of the projected t-J model}\label{ap:NoSOC-BCS}
\begin{figure}[h]
  \setlength{\unitlength}{1mm}

   \includegraphics[width = 0.45\textwidth]{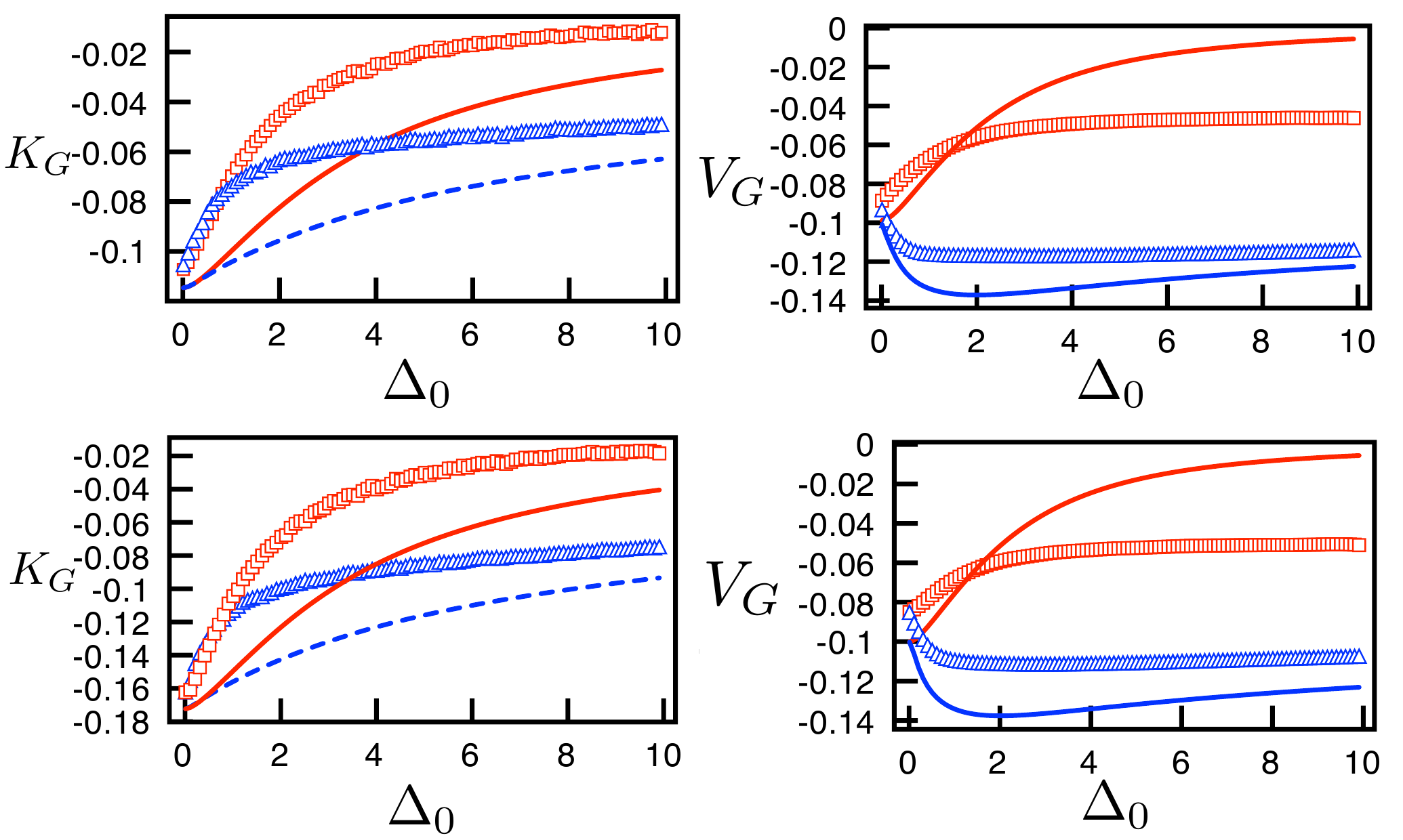}
\caption{{\small
Comparison of Monte Carlo and Gutzwiller approximation results for the kinetic energy $K_G$ and magnetic energy $V_G$. On the left we compare exact Monte Carlo results for the kinetic energy to the approximation $K_G$ while on the right we explore the comparison of $V_G$ to Monte Carlo results. All plots
are for $N=82$ lattice sites, the top plots are for a system with $N_H=4$ holes while the bottom plots are $N_H=8$
holes. At each doping level we have tried two different order parameter symmetries, $d$-wave and $s$-wave. For
$d$-wave symmetry with $\Delta_\kv = \Delta_0(\cos(k_x)-\cos(k_y))$ we have plotted the Gutzwiller approximations
given in Eqx.~(\ref{Kgut}) and (\ref{defVg3})   using a dashed line (blue online) and numerical Monte Carlo results using triangles
(blue online). For $s$-wave symmetry we have $\Delta_\kv = \Delta_0$ and have plotted the Gutzwiller
approximation using a solid line (red online) and numerical Monte Carlo results using squares (red online). In
both figures we have set $t=1$ and $\xi_\kv = \epsilon_\kv -\mu$ where $\epsilon_\kv$ is defined in the text and
$\mu$ is the chemical potential.  We have also scaled the kinetic energy by the number of lattice sites, $N_s$.
     }
     }\label{fig:compare}
\end{figure}

This appendix will present detailed results for the $A,M,B=0$ limit of the Gutzwiller approximation developed in Section \ref{sec:Gutwiller}. We begin by recalling the form of the BCS wavefunction in its most generic form. We begin with
ground state
\begin{equation}
|\psi_{BCS}\rangle = \prod_{\kv} (u_\kv+v_{\kv}c^\dagger_{\kv,\uparrow} c^\dagger_{-\kv,\downarrow})|0\rangle
\end{equation}
where $u_\kv = \sqrt{\frac{E_\kv+\xi_\kv}{2E_\kv}}$ and $v_\kv =
e^{i\phi_\kv}\sqrt{\frac{E_\kv-\xi_\kv}{2E_\kv}}$ where $E_\kv = \sqrt{\xi_\kv^2 +|\Delta_{\kv}|^2}$ and
$\phi_\kv$ is the phase of $\Delta_{\kv}$. In this context the functions $\Delta_\kv$ and $\xi_\kv$ are
variational parameters we are interested in optimizing. In order to explore how well our Gutzwiller approximation for the energies $K_G$ and $V_G$ does in comparison to numerical results for the full expectation values in the state $P_G|\psi_{Var}\rangle$ we have carried out Monte Carlo calculations to compare to $K_G$ and $V_G$. Our results at doping values of $x\simeq 0.04878$ and $x\simeq 0.09756$ for $d$-wave and $s$-wave order parameter symmetries are presented in Figure \ref{fig:compare}. Following the case made in Ref. [\onlinecite{zhang}], we contend that the Gutzwiller approximation does a fair job at reproducing the quantitative values of $K_G$ and $V_G$ and an excellent job at
reproducing the qualitative behavior of the two as it gets the general trend of the curve correct.

With some credibility for the Gutzwiller approximation established we now use $|\psi_{BCS}\rangle$ to evaluate a variational energy which we can minimize to find $\xi_\kv$ and $\Delta_\kv$. Calculating the appropriate expectation values the Gutzwiller approximated variational energy reads
\begin{eqnarray}
E(\Delta_\kv,\xi_\kv) &=& -\sum_{\kv} (g_t\epsilon_\kv-\mu) \left(\frac{\xi_\kv}{E_\kv}\right)
\\ \nonumber &-&\frac{3\tilde{J}}{8N} (d_x^2+d_y^2+2e_x^2+2e_y^2)
\end{eqnarray}
where $\epsilon_\kv = -2t(\cos{k_x}+\cos{k_y})$ is the tight binding spectrum, $E_{\kv}=\sqrt{\xi_{\kv}^2+\Delta_{\kv}^2}$ and we have defined $d_{i} = \sum_{\kv} \left(\frac{\Delta_\kv \cos(k_i)}{E_\kv}\right)$, $e_{i} = \sum_{\kv}
\left(\frac{\xi_\kv \cos(k_i)}{E_\kv}\right)$ and $\mu$ is determined self-consistently through $x =
\frac{1}{N}\sum_\kv\left(\frac{\xi_\kv }{E_\kv}\right)$.  Minimizing the above functional with respect to
$\Delta_\kv$ and $\xi_\kv$ leads to the following results
\begin{eqnarray}
&&\Delta_\kv = \Delta_x \cos(k_x)+\Delta_y \cos(k_y) \\ \nonumber &&\xi_\kv =g_t\epsilon_\kv -\mu + \xi_x
\cos(k_x)+\xi_y \cos(k_y) \\ \nonumber
\end{eqnarray}
where the parameters $\Delta_i$ and $\xi_i$ must be determined self-consistently through the solutions to
following equations
\begin{eqnarray}\label{SelfCons}
&&\Delta_i = \frac{3g_JJ}{4N}\sum_{\kv} \left(\frac{\Delta_\kv \cos(k_i)}{E_\kv}\right)  \\ \nonumber
&&\xi_i=\frac{3g_JJ}{4N}\sum_{\kv} \left(\frac{\xi_\kv \cos(k_i)}{E_\kv}\right)  \\ \nonumber
\end{eqnarray}

\begin{figure}[h]
  \setlength{\unitlength}{1mm}

   \includegraphics[width = 0.4\textwidth]{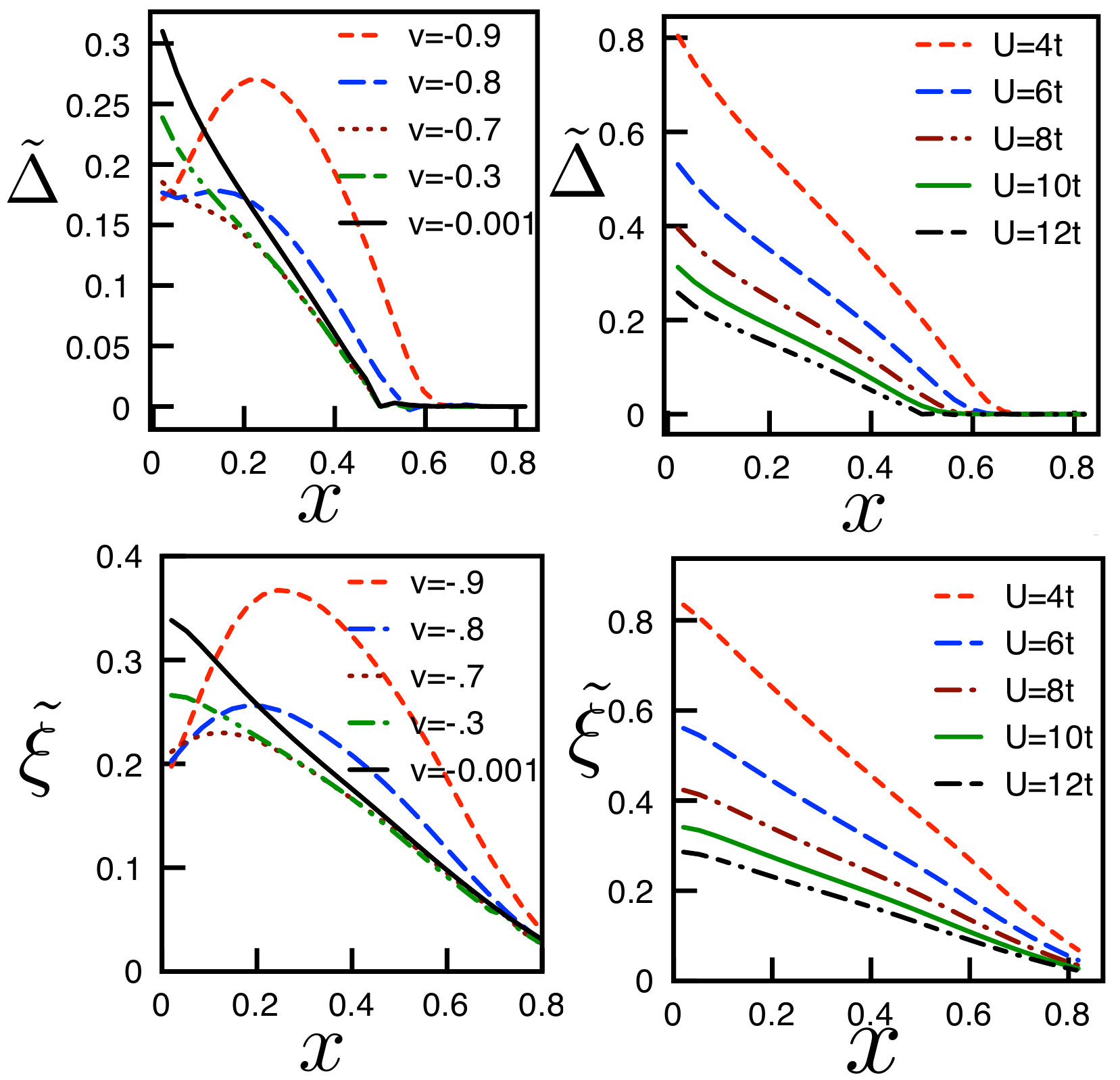}
\caption{{\small
Results of self-consistent calculations for varying values of $V$ and $U$. The top left figure shows the results for
$\tilde{\Delta}$ while the bottom left one shows $\tilde{\xi}$. For these left figures we have fixed $U=12t$ and allowed $V$ to vary. On the right,  the top figure shows the results for
$\tilde{\Delta}$ while the bottom one shows $\tilde{\xi}$ when $V$ is fixed to $V=-0.2U$ and $U$ is allowed to vary. For the economy of space in the legend of this figure we have defined $v=V/U$.
    }}\label{fig:OP}
\end{figure}

 We have solved the self consistent equations in Eq.~(\ref{SelfCons}) over a space of $x$, $U$ and $V$ values.
 We have fixed $U=12t$ as is appropriate from studies of the Hubbard model in the past\cite{zhang, Paramekanti1}. We
 have found that the solution is that of a $d$-wave superconductor, $\Delta_x=-\Delta_y$ with
 $|\Delta_x|=|\Delta_y|=\tilde{\Delta}$, and $\xi_x=\xi_y=-|\tilde{\xi}|$. The results of our calculations are
 presented in Fig. \ref{fig:OP}.

To close this appendix we will present calculations of the superconducting order parameter. The parameter
$\tilde{\Delta}$ is not exactly the ``order parameter" as we must take into account the Gutzwiller projection
operator. We define the superconducting order parameter as follows
\begin{equation}
\Delta_{SC}(i,j) = \frac{\langle \psi_{BCS} |P_G\left( c_{i,\uparrow}^\dagger
c_{j,\downarrow}^\dagger-c_{i,\downarrow}^\dagger c_{j,\uparrow}^\dagger\right)P_G|\psi_{BCS}\rangle}{\langle
\psi_{BCS} |P_G|\psi_{BCS}\rangle}
\end{equation}
The Gutzwiller approximation for this quantity is then as follows
 \begin{equation}
\Delta_{SC}(i,j) \simeq g_{\Delta} \frac{\langle \psi_{BCS} |\left( c_{i,\uparrow}^\dagger
c_{j,\downarrow}^\dagger-c_{i,\downarrow}^\dagger c_{j,\uparrow}^\dagger\right)|\psi_{BCS}\rangle}{\langle
\psi_{BCS} |\psi_{BCS}\rangle}
\end{equation}
 where $g_{\Delta}=2x/(1+x)$ again comes from a proper weighting using relative probabilities. Upon writing the
 above in momentum space one can easily show the following result for nearest neighbor $(i,j)$.
   \begin{equation}\label{OPi}
\Delta_{SC}(i) =\frac{ 4g_{\Delta}}{3 \tilde{J}} \Delta_{i}
\end{equation}

\begin{figure}[h]
   \includegraphics[width = 0.4\textwidth]{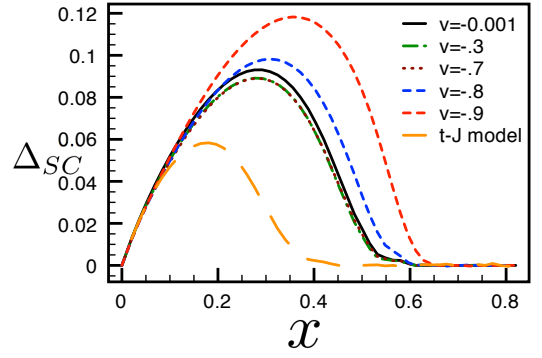}
\caption{{\small
Plot of order parameter as given in Eq.~(\ref{OPi}) at various values of  $v=V/U$ with $U=12t$. For comparison we
have also plotted the value of $\Delta_{SC}$ obtained from the traditional $t-J$ model.
     }
     }\label{fig:SCOP}

     \section{Spin Renormalization Factors at Finite Spin-Orbit Coupling} \label{sec:Gfactors}
\end{figure}

We have plotted our results for this quantity at the same values of $U$ as on the left of Fig.~\ref{fig:OP} and have compared our results to those of the bare $t-J$ model. The results are in Figure  \ref{fig:SCOP}. We see that the order parameter $\Delta_{SC}$ is a non-monotonic function of of the doping $x$ which grows from zero to
a maximum and then returns to zero at $x_c$. Both the maximum value of $\Delta_{SC}$ and the $x$ value at which
it is achieved vary only slightly with $v$ and for large $v$ both tend to increase with $v$. We see that our analysis here leads to an enhancement of $\Delta_{SC}$. This makes physical sense as the
difference between the model considered here and the $t-J$ model is an attractive nearest neighbor interaction,
something that should favour $d$-wave superconductivity.

     Here we discuss how to perform the Gutzwiller approximation for the term $\sum_{i,\delta} \tilde{J}_{\delta}^{\mu\nu} \Sv_{i}^\mu\Sv_{i+\delta}^\nu$ where $\tilde{J}_{\delta}^{\mu\nu} $ is given in Eq. (\ref{JGA1})\cite{farrellthesis}. We begin by dividing this term into four separate terms given by
\begin{eqnarray} \label{exterms}
H_{J,\text{i}}&=&\frac{J_1-J_4}{4} \sum_{i,\delta} \left(S_{i+\delta}^{+} S_{i}^{-} +S_{i+\delta}^{-} S_{i}^{+} \right)\\ \nonumber
H_{J,\text{ii}}&=&\frac{J_1-J_2+J_4}{2} \sum_{i,\delta}  S_{i+\delta}^{z} S_{i}^{z}\\ \nonumber
H_{J,\text{iii}}&=&\frac{J_2}{4} \sum_{i,\delta}a(\delta) ( S_{i+\delta}^{+} S_{i}^{+} +S_{i+\delta}^{-} S_{i}^{-})  \\ \nonumber
H_{J,\text{iv}}&=&\frac{J_3}{2} \sum_{i,\delta}\left(  \delta \cdot (\hat{y}+i \hat{x})S_{i+\delta}^{z} S_{i}^{+}+ \delta \cdot (\hat{y}-i \hat{x})S_{i+\delta}^{z} S_{i}^{-}\right)
\end{eqnarray}
where we have defined the spin raising and lowering operators $S^+_{i} = c_{i,\uparrow}^\dagger c_{i,\downarrow}$ and $S^-_{i} = c_{i,\downarrow}^\dagger c_{i,\uparrow}$. Let us discuss the renormalization of each term above in the order listed.

$H_{J,\text{i}}$ takes two occupied nearest neighbour sites and flips their spins. We discuss $S_{i+\delta}^{+} S_{i}^{-}$ for concreteness while keeping in mind that this discussion immediately extends to the second term. In a Gutzwiller projected state we only require that sites $i$ and $i+\delta$ are occupied  by an up spin and a down spin respectively, while in a non-projected state we not only require states $(i,\uparrow)$ and $(i+\delta, \downarrow)$ to be occupied but also need $(i,\downarrow)$ and $(i+\delta, \uparrow)$ to be empty. These considerations lead to the renormalization factor
  \beq
  g_{J,\text{i}} = \frac{1}{(1-\langle n_{\uparrow} \rangle)(1-\langle n_{\downarrow}\rangle)}
  \eeq
  and we make the approximation $H_{J,\text{i}} \to   g_{J,\text{i}} H_{J,\text{i}}$.

  Moving on, $H_{J,\text{ii}}$ is relatively straightforward to understand, it measures the $z$-projection of spin on two nearest neighbour sites . Owing to the fact that we have made the approximation that the number of electrons on a given lattice site is the same in both the projected and unprojected state and that $S_{i}^{z}= n_{i, \uparrow}-n_{i, \downarrow}$ we obtain equal weighting in both states and so $g_{J, \text{ii}}=1$.

  Third, we have $H_{J,\text{iii}}$ which either flips two neighbouring down spins to up spins or flips two neighbouring up spins to down spins. We discuss the latter for concreteness (symbolically this corresponds to the  $S_{i+\delta}^{+} S_{i}^{+} $ term). Naively,  in a Gutzwiller projected state we only require that sites $i$ and $i+\delta$ are both occupied  by a down spin, while in a non-projected state we not only require states $(i,\downarrow)$ and $(i+\delta, \downarrow)$  to be occupied but also need  $(i,\uparrow)$ and $(i+\delta, \uparrow)$ to be empty. This would lead to the factor $  g_{J,\text{iii}} = \frac{1}{(1-\langle n_{\downarrow} \rangle)(1-\langle n_{\downarrow}\rangle)}$ and ultimately to a renormalized $H_{J,\text{iii}}$ term which is not Hermitian for general values of $\langle n_{\downarrow} \rangle$ and $\langle n_{\uparrow} \rangle$. It turns out that to remedy this problem we must form a geometric average between  the probability for the forward process $S_{i+\delta}^{+} S_{i}^{+} $ and the probability for it to happen in reverse \cite{Ko} ({\it i.e.} for two up spins to flip to two down spins) . This leads to a geometric average between $\frac{1}{(1-\langle n_{\downarrow} \rangle)(1-\langle n_{\downarrow}\rangle)}$ and $\frac{1}{(1-\langle n_{\uparrow} \rangle)(1-\langle n_{\uparrow}\rangle)}$ and ultimately to the renormalization factor
  \beq
   g_{J,\text{iii}} =  \frac{1}{(1-\langle n_{\uparrow}\rangle)(1-\langle n_{\downarrow}\rangle)}
  \eeq

  Finally we discuss the renormalization of $H_{J,\text{iv}}$. We discuss $S_{i+\delta}^{z} S_{i}^{+}$ for concreteness. This term flips an up spin to a down spin on site $i$ and then measures the value of the $z$-projection of spin on a nearest neighbour site. In the Gutzwiller projected state we require only that state $(i, \downarrow)$ be occupied while in the unprojected state we need  $(i, \downarrow)$ and  $(i, \uparrow)$ empty. Again we need to be careful with the equal weighting of forward and backwards processes\cite{Ko}. Taking the reverse process into account then leads to the final renormalization factor
  \begin{eqnarray}
      g_{J,\text{iv}} &=&  \frac{1}{\sqrt{(1-\langle n_{\uparrow}\rangle)(1-\langle n_{\downarrow}\rangle)}}
 \end{eqnarray}

\end{document}